\documentclass[article,shortnames,nojss]{jss}

\usepackage{thumbpdf,lmodern}

\usepackage{framed} 
\usepackage{lstpython} 
\usepackage{natbib} 
\usepackage{xspace} 
\usepackage{booktabs}
\usepackage{multirow}

\usepackage{amsmath,amssymb,amsthm}
\usepackage{physics}
\usepackage{bm}

\theoremstyle{definition}
\newtheorem*{remark}{Remark}

\usepackage{algorithm}
\usepackage[noend]{algpseudocode}

\usepackage{array}
\usepackage{makecell}
\newcolumntype{L}[1]{>{\raggedright\let\newline\\\arraybackslash\hspace{0pt}}p{#1}}
\newcolumntype{C}[1]{>{\centering\let\newline\\\arraybackslash\hspace{0pt}}p{#1}}
\newcolumntype{R}[1]{>{\raggedleft\let\newline\\\arraybackslash\hspace{0pt}}p{#1}}

\usepackage[normalem]{ulem}
\newcommand{\mcode}[1]{\textnormal{\code{#1}}}

\newcommand{\fct}[1]{\code{#1()}}

\makeatletter
\newcommand{\iid}{\stackrel {\operator@font{iid}}{\sim}}
\newcommand{\ind}{\stackrel {\operator@font{ind}}{\sim}}
\makeatother
\DeclareMathOperator{\N}{Normal}

\DeclareMathOperator*{\argmax}{arg\,max}
\renewcommand{\L}{\textrm{lin}}

\let\var\relax
\DeclareMathOperator{\var}{var}

\DeclareMathOperator{\diag}{diag}

\renewcommand{\aa}{{\bm a}}
\renewcommand{\AA}{{\bm A}}
\newcommand{\bb}{{\bm b}}
\newcommand{\cc}{{\bm c}}
\newcommand{\ff}{{\bm f}}
\renewcommand{\gg}{{\bm g}}

\newcommand{\yy}{{\bm y}}
\newcommand{\xx}{{\bm x}}
\newcommand{\vv}{{\bm v}}

\newcommand{\zz}{{\bm z}}
\renewcommand{\AA}{{\bm A}}
\newcommand{\BB}{{\bm B}}
\newcommand{\CC}{{\bm C}}
\newcommand{\DD}{{\bm D}}
\newcommand{\JJ}{{\bm J}}
\newcommand{\QQ}{{\bm Q}}
\newcommand{\RR}{{\bm R}}
\newcommand{\UU}{{\bm U}}
\newcommand{\VV}{{\bm V}}
\newcommand{\WW}{{\bm W}}
\newcommand{\XX}{{\bm X}}
\newcommand{\YY}{{\bm Y}}
\newcommand{\ZZ}{{\bm Z}}
\newcommand{\eps}{{\bm \epsilon}}
\newcommand{\eet}{{\bm \eta}}

\newcommand{\tth}{{\bm \theta}}
\newcommand{\pph}{{\bm \phi}}
\newcommand{\lla}{{\bm \lambda}}

\newcommand{\mmu}{{\bm \mu}}
\newcommand{\ssi}{{\bm \sigma}}

\newcommand{\OOm}{{\bm \Omega}}
\newcommand{\SSi}{{\bm \Sigma}}
\newcommand{\TTh}{{\bm \Theta}}
\newcommand{\bz}{{\bm 0}}
\newcommand{\Id}{{\bm I}}
\newcommand{\dt}{\Delta t}

\newcommand{\bO}{\mathcal{O}}
\newcommand{\mi}{\zz}
\newcommand{\vi}{\VV}
\newcommand{\Ell}{{\mathcal{L}}}
\newcommand{\tmin}{t_{\mathrm{min}}}
\newcommand{\tmax}{t_{\mathrm{max}}}
\newcommand{\xobs}{{\bm{s}}}
\newcommand{\xun}{{\bm{u}}}
\newcommand{\hxobs}{\hat{\xobs}}
\newcommand{\ipar}{{\bm{\mathcal{I}}}}
\newcommand{\upar}{{\bm{\Psi}}}
\newcommand{\utrans}{{\bm{\mathcal{G}}}}

\newcommand{\kpredict}{\mcode{kalman\_predict}}
\newcommand{\kforecast}{\mcode{kalman\_forecast}}
\newcommand{\kupdate}{\mcode{kalman\_update}}
\newcommand{\ksample}{\mcode{kalman\_sample}}
\newcommand{\ksmooth}{\mcode{kalman\_smooth}}
\newcommand{\kcond}{\mcode{kalman\_cond}}
\newcommand{\inter}{\mcode{interrogate}}
\newcommand{\kbpredict}{\mcode{kalman\_block\_predict}}
\newcommand{\kbforecast}{\mcode{kalman\_block\_forecast}}
\newcommand{\kbupdate}{\mcode{kalman\_block\_update}}
\newcommand{\kbsample}{\mcode{kalman\_block\_sample}}
\newcommand{\kbsmooth}{\mcode{kalman\_block\_smooth}}
\newcommand{\kbcond}{\mcode{kalman\_block\_cond}}
\newcommand{\binter}{\mcode{block\_interrogate}}
\newcommand{\nlogpdf}{\mcode{normal\_logpdf}}

\newcommand{\curr}{{(\textnormal{curr})}}
\newcommand{\prop}{{(\textnormal{prop})}}

\algnewcommand{\LeftComment}[1]{\Statex \(\triangleright\) #1}

\newcommand{\jax}{\pkg{JAX}\xspace}
\newcommand{\python}{\proglang{Python}\xspace}
\newcommand{\matlab}{\proglang{Matlab}\xspace}
\newcommand{\rlang}{\proglang{R}\xspace}
\newcommand{\julia}{\proglang{Julia}\xspace}
\newcommand{\rodeo}{\pkg{rodeo}\xspace}
\newcommand{\diffrax}{\pkg{diffrax}\xspace}

\newcommand{\fenrir}{\textnormal{Fenrir}\xspace}
\newcommand{\dalton}{\textnormal{DALTON}\xspace}
\newcommand{\chkrebtii}{\textnormal{Chkrebtii}\xspace}
\newcommand{\magi}{\textnormal{MAGI}\xspace}
\newcommand{\bjax}{\pkg{BlackJAX}\xspace}

\author{Mohan Wu\\University of Waterloo
  \And Martin Lysy\\University of Waterloo}
\Plainauthor{Mohan Wu, Martin Lysy}

\title{\rodeo: Probabilistic Methods of Parameter Inference for Ordinary Differential Equations}
\Plaintitle{rodeo: Probabilistic Methods of Parameter Inference for Ordinary Differential Equations}
\Shorttitle{\rodeo: Probabilistic Methods of ODE Parameter Inference}

\Abstract{
  Parameter estimation for ordinary differential equations (ODEs) plays a fundamental role in the analysis of dynamical systems.  Generally lacking closed-form solutions, ODEs are traditionally approximated using deterministic solvers.  However, there is a growing body of evidence to suggest that probabilistic ODE solvers produce more reliable parameter estimates by better accounting for numerical uncertainty.
  Here we present \rodeo, a \python library providing a fast, lightweight, and extensible interface to a broad class of probabilistic ODE solvers, along with several associated methods for parameter inference.  At its core, \rodeo provides a probabilistic solver that scales linearly in both the number of evaluation points and system variables.   Furthermore, by leveraging state-of-the-art automatic differentiation (AD) and just-in-time (JIT) compiling techniques, \rodeo is shown across several examples to provide fast, accurate, and scalable parameter inference for a variety of ODE systems.
}

\Plainkeywords{Differential equations, probabilistic solution, Kalman filter, Python, JAX}

\Address{
  Mohan Wu, Martin Lysy\\
  Department of Statistics and Actuarial Science\\
  University of Waterloo\\
  E-mail: \email{mlysy@uwaterloo.ca}
}

\begin{document}


\section{Introduction}
Parameter estimation for ordinary differential equations (ODEs) is an important problem in the natural sciences and engineering.  
Since most ODEs do not have closed-form solutions, they must be approximated by numerical methods. Traditionally, this has been done with deterministic algorithms~\citep[e.g.,][]{butcher08, griffiths.higham10, atkinson.etal09}. However, an emerging field of \emph{probabilistic numerics}~\citep{diaconis88, skilling92, hennig.etal15} indicates that probabilistic ODE solvers, which directly account for uncertainty in the numerical approximation, provide more reliable parameter estimates in ODE learning problems~\citep{chkrebtii.etal16,conrad.etal17}.

%

\begin{table}[H]
    \caption{Various probabilistic ODE solvers.  \textsl{Method/Software:} The proposed method in the accompanying reference(s) or the name of the software library if it is publicly available. \textsl{Latent Variables:} Whether the proposed method involves a large number of latent variables which need to be integrated out using computationally intensive methods. \textsl{Implementation:} The programming language employed if an implementation is available.  \textsl{Compiled:} Whether the implementation is written in compiled code.  \textsl{Autodiff:} Whether the implementation enables automatic differentiation of user-defined models.
}\label{tab:solvers}
    \centering
    \scalebox{0.9}{
    \begin{tabular}{L{8cm}C{2cm}ccc}
        \toprule
        Method/Software & Latent Variables & Implementation & Compiled & Autodiff \\
        \midrule
        \makecell[l]{Gaussian Process Regression \\ \citep{calderhead.etal09}} & Y & \matlab & N & N \\ \hline
        \makecell[l]{Adaptive Gradient Matching \\ \citep{dondelinger.etal13}} & Y &  &  &  \\ \hline
        \makecell[l]{\pkg{odegp} \\ \citep{barber.wang14}} & Y & \matlab & N & N \\ \hline
        \makecell[l]{Gaussian Markov Runge-Kutta \\ \citep{schober.etal14}} & Y & \matlab & N & N \\ \hline
        \makecell[l]{\pkg{uqdes} \\ \citep{chkrebtii.etal16}} & Y & \matlab & N & N \\ \hline
        \makecell[l]{Bayesian Quadrature Filtering \\ \citep{kersting.hennig16}} & Y &  &  &  \\ \hline
        \makecell[l]{Adams-Bashforth \\ \citep{teymur.etal16}} & Y &  &  &  \\ \hline
        \makecell[l]{\pkg{CollocInfer} \\ \citep{hooker.etal16}} & Y & \rlang/\matlab & N & N \\ \hline
        \makecell[l]{\pkg{deBInfer} \\ \citep{supan.etal17}} & Y & \rlang & N & N \\ \hline
        \makecell[l]{\pkg{GPmat} \\ \citep{ghosh.etal17}} & Y & \matlab & N & N \\ \hline
        \makecell[l]{Scalable Variational Inference \\ \citep{gorbach.etal17}} & Y &  &  &  \\ \hline
        \makecell[l]{Multiphase MCMC \\ \citep{lazarus.etal18}} & Y &  &  &  \\ \hline
        \makecell[l]{\pkg{FGPGM} \\ \citep{wenk.etal19}} & Y & \python & N & N \\ \hline
        \makecell[l]{\pkg{deGradInfer} \\ \citep{macdonal.dondelinger20}} & Y & \rlang & N & N \\ \hline
        \makecell[l]{\pkg{simode} \\ \citep{dattner.yaari20}} & Y & \rlang & N & N \\ \hline
        \makecell[l]{\pkg{KGode} \\ \citep{niu.etal21}} & Y & \rlang & N & N \\ \hline
        \makecell[l]{\pkg{MAGI} \\ \citep{yang.etal21}} & Y & \rlang/\python/\matlab & Y & N \\ \hline
        \makecell[{{L{8cm}}}]{\pkg{ProbNum} \\ \citep{tronarp.etal18, schober.etal19, kramer20, wenger21, kramer21}} & N & \python & N & N \\ \hline
        \makecell[l]{\pkg{pCODE} \\ \citep{wang.cao22}} & Y & \rlang & N & N \\ \hline
        \makecell[{{L{8cm}}}]{\pkg{ProbNumDiffEq.jl}/\pkg{ProbDiffEq} \\ \citep{bosch21, bosch22, tronarp22, kramer24}} & N & \julia/\python & Y & Y \\ \hline
        \bottomrule
    \end{tabular}}
\end{table}
A number of recently proposed probabilistic ODE solvers are summarized in Table~\ref{tab:solvers}.  
Of these methods, 
few have been compared directly to deterministic solvers in terms of speed and accuracy. Partly, this is because many probabilistic solvers incur a high cost per time step, e.g., involving large matrix manipulations, or having a large number of latent variables to integrate out via computationally intensive methods such as Markov chain Monte Carlo (MCMC) or particle filtering (see the ``Latent Variables'' column of Table~\ref{tab:solvers}).  
It is also because in publicly available software implementations, design considerations are often guided by simplicity of interface rather than performance optimization (see ``Compiled'' column of Table~\ref{tab:solvers}).

A notable exception to these limitations is the \julia library \pkg{ProbNumDiffEq.jl}~\citep{bosch21,bosch22, tronarp22},
which has been shown to outperform a number of deterministic solvers~\citep{bosch22}.  Moreover, \pkg{ProbNumDiffEq.jl} is seamlessly compatible with the \julia automatic differentiation (AD) and just-in-time (JIT) compilation engines,
which allow for fast deployment of many gradient-based parameter inference algorithms (more on this in Section~\ref{sec:examples}).

Here we present \rodeo\footnote{https://github.com/mlysy/rodeo}, a fast and lightweight \python library providing several probabilistic methods of ODE parameter inference.  Underlying these methods is a nonlinear Bayesian filtering paradigm which \rodeo uses to implement a broad class of probabilistic ODE solvers~\citep{chkrebtii.etal16,tronarp.etal18,schober.etal19,yang.etal21}.  Built on top of the \jax linear algebra library~\citep{jax2018github}, the \rodeo algorithms are written in AD-compatible, JIT-compiled code in the widely used \pkg{NumPy} syntax.  By default, these algorithms are implemented to scale linearly in both the number of discretization points and ODE system variables~\citep{kramer21}.

In several respects, the \rodeo probabilistic ODE solver is similar to that in \pkg{ProbDiffEq}~\citep{kramer24}, a \python port of a subset of \pkg{ProbNumDiffEq.jl} also written in \pkg{JAX}.  The \pkg{ProbDiffEq} ODE solver has several features which \rodeo currently lacks, such as adaptive step-size selection and filter preconditioning.  However,
while \pkg{ProbDiffEq} focuses primarily on providing a fast probabilistic ODE solver, the main focus of \rodeo is to offer a wide range of ODE parameter inference methods within the Bayesian nonlinear filtering framework, integrating approaches developed by several research groups~\citep[e.g.,][]{chkrebtii.etal16, tronarp.etal18, tronarp22, yang.etal21, wu.etal24}.
Several examples illustrate the speed and flexibility of \rodeo, allowing users to choose from a collection of built-in inference methods depending on the problem setting.



The rest of this paper is organized as follows. Section~\ref{sec:method} reviews probabilistic ODE solvers using the Bayesian filtering paradigm. Section~\ref{sec:parinf} introduces the parameter inference methods available in \rodeo. Section~\ref{sec:packstruct} discusses some design considerations and implementation details.  Section~\ref{sec:examples} demonstrates the \rodeo application programming interface (API) through several examples, showcasing each built-in inference method. Section~\ref{sec:disc} concludes with a discussion directions for future research.

\newpage

\section{Probabilistic ODE Solvers via Bayesian Filtering}\label{sec:method}

\rodeo
is designed to solve arbitrary-order multivariable ODE systems which satisfy an initial value problem (IVP).  For a univariate function $x(t)$, an ODE-IVP is of the form
\begin{equation}
    \label{eq:hode}
    \WW\xx(t) = \ff(\xx(t), t), \qquad \xx(\tmin) = \vv, \qquad t \in [\tmin, \tmax],
\end{equation}
where $\xx(t) = \big(x^{(0)}(t), x^{(1)}(t), \ldots, x^{(p-1)}(t)\big)$ consists of $x(t) = x^{(0)}(t)$ and its first $p-1$ derivatives, $\WW_{r \times p}$ is a coefficient matrix, and $\ff(\xx(t), t) = \big(f_1(\xx(t), t), \ldots, f_r(\xx(t), t)\big)$ is a nonlinear function representing $r$ equations.  The ODE in~\eqref{eq:hode} can always be written in the more usual ``first-order'' form,
\begin{equation}
  \label{eq:ode1}
  \dv{t} x^{(i)}(t) = g_i(\xx(t), t), \qquad i=0,\ldots,p-2.
\end{equation}
Equivalently, any ODE in the first-order form~\eqref{eq:ode1} can be written in the form of~\eqref{eq:hode} with $\ff(\xx(t), t) = (g_0(\xx(t), t), \ldots, g_{p-2}(\xx(t), t))$ and $\WW_{(p-1)\times p} = [\bz_{(p-1)\times 1} \mid \Id_{(p-1) \times (p-1)}]$.  While the first-order form~\eqref{eq:ode1} is unique, the formulation in~\eqref{eq:hode} 
often allows a given ODE-IVP to be specified in multiple ways, which in turn could affect the accuracy of the stochastic solver to be described below. 

Unlike deterministic solvers, \rodeo employs a probabilistic approach to solving~\eqref{eq:hode} based on a widely adopted paradigm of Bayesian nonlinear filtering~\citep{tronarp.etal18}.  This approach consists of putting a Gaussian Markov process prior on $\xx(t)$, and updating it with information from the ODE-IVP~\eqref{eq:hode} at time points $t = t_0, \ldots, t_N$, where $t_n = n(\tmax-\tmin)/N$.  Specifically, let $\xx_n = \xx(t_n)$ and consider the general indexing notation $\xx_{m:n} = (\xx_m, \ldots, \xx_n)$.  If $\xx(t)$ is the solution to~\eqref{eq:hode}, we would have $\zz_n = \WW \xx_n - \ff(\xx_n, t_n) = \bz$.  Based on this observation,  \cite{tronarp.etal18} consider a state-space model in $\xx_n$ and $\zz_n$ of the form
\begin{equation}\label{eq:bnf}
    \begin{aligned}
        \xx_{n+1} \mid \xx_n & \sim \N(\QQ \xx_n, \RR) \\
        \mi_n & \ind \N(\WW \xx_n - \ff(\xx_n, t_n), \vi_n),
    \end{aligned}
\end{equation}
where $\xx_0 = \vv$, $\QQ$ and $\RR$ are determined by the Gaussian Markov process prior, and $\vi_{1:N}$ are tuning parameters.  The stochastic ODE solution is then given by the posterior distribution 
\begin{equation}\label{eq:solpost}
  p(\xx_{1:N} \mid \mi_{1:N} = \bz).
\end{equation}

As $N \to \infty$, the stochastic ODE solution~\eqref{eq:solpost} gets arbitrarily close to the true ODE solution as the computational complexity increases~\citep{kersting20}.  
However, the posterior distribution~\eqref{eq:solpost} generally
cannot be sampled from directly.  Alternatives include Markov chain
Monte Carlo sampling~\citep{chkrebtii.etal16,yang.etal21} and particle
filtering~\citep{tronarp.etal18}.  A less accurate, but ostensibly much
faster approach is to linearize~\eqref{eq:bnf} using Taylor
expansions~\citep{tronarp.etal18}.  This amounts to the surrogate model
\begin{equation}\label{eq:mi}
    \begin{aligned}
        \xx_{n+1} \mid \xx_n & \sim \N(\QQ \xx_n, \RR) \\
        \mi_n & \ind \N((\WW + \BB_n)\xx_n + \aa_n, \vi_n),
    \end{aligned}
\end{equation}
where $\aa_n, \BB_n$ and $\vi_n$
are obtained sequentially by a process called \emph{model interrogation}~\citep{chkrebtii.etal16} (see Section~\ref{sec:mi}).  The upshot is that a random draw from the posterior
\begin{equation}\label{eq:plin}
  p_\L(\xx_{1:N} \mid \mi_{1:N} = \bz)
\end{equation}
induced by the working model~\eqref{eq:mi} can be easily obtained using the Kalman filtering and smoothing algorithms.  Denote the posterior mean and variance of the solution process under the working model~\eqref{eq:mi} as
\begin{equation}
  \begin{aligned}
    \mmu_{m|n} & = \E_\L[\xx_m \mid \mi_{1:n} = \bz], \\
    \SSi_{m|n} & = \var_\L(\xx_m \mid \mi_{1:n} = \bz).
  \end{aligned}
\end{equation}
Using this notation, the steps of the Kalman ODE solver are presented in Algorithm~\ref{alg:Kalman}. The formulation of the Kalman recursions used in Algorithm~\ref{alg:Kalman} can be found in Appendix~\ref{sec:kalmanfun}.
\begin{algorithm}[!htb]
  \caption{The Kalman ODE solver.}\label{alg:Kalman}
  \begin{algorithmic}[1]
    \Procedure{\code{kalman\_ode\_sim}}{$\WW, \ff(\xx, t), \vv, \QQ, \RR$}
    \State $\mmu_{0|0}, \ldots, \mmu_{N|N}, \SSi_{0|0}, \ldots, \SSi_{N|N} \gets \mcode{kalman\_ode\_forward}(\WW, \ff(\xx, t), \vv, \QQ, \RR)$
    \State $\xx_N \sim \N(\mmu_{N|N}, \SSi_{N|N})$
    \For{$n=N-1:0$}
    \State $\xx_n \gets \ksample(\xx_{n+1}, \mmu_{n|n}, \SSi_{n|n}, \mmu_{n+1|n}, \SSi_{n+1|n}, \QQ)$
    \EndFor
    \State \textbf{return} $\xx_{0:N}$
    \EndProcedure
    \State
    \Procedure{\code{kalman\_ode\_mv}}{$\WW, \ff(\xx, t), \vv, \QQ, \RR$}
    \State $\mmu_{0|0}, \ldots, \mmu_{N|N}, \SSi_{0|0}, \ldots, \SSi_{N|N} \gets \mcode{kalman\_ode\_forward}(\WW, \ff(\xx, t), \vv, \QQ, \RR)$
    \For{$n=N-1:0$}
    \State $\mmu_{n|N}, \SSi_{n|N} \gets \ksmooth(\mmu_{n+1|N}, \SSi_{n+1|N}, \mmu_{n|n}, \SSi_{n|n}, \mmu_{n+1|n}, \SSi_{n+1|n}, \QQ)$
    \EndFor
    \State \textbf{return} $\mmu_{0:N|N}, \SSi_{0:N|N}$
    \EndProcedure
    \State
    \Procedure{\code{kalman\_ode\_forward}}{$\WW, \ff(\xx, t), \vv, \QQ, \RR$}
    \State $\mmu_{0|0}, \Sigma_{0|0} \gets \vv, \bz$
    \State $\mi_{1:N} \gets \bz$
    \For{$n=1:N$}
    \State $\mmu_{n|n-1}, \SSi_{n|n-1} \gets \kpredict(\mmu_{n-1|n-1}, \SSi_{n-1|n-1}, \bz, \QQ, \RR)$
    \State $\aa_n, \BB_n, \vi_n \gets \inter(\mmu_{n|n-1}, \SSi_{n|n-1}, \WW, \ff(\xx, t))$\label{ln:inter} \Comment{See Table~\ref{tab:inter}}
    \State $\mmu_{n|n}, \SSi_{n|n} \gets \kupdate(\mmu_{n|n-1}, \SSi_{n|n-1}, \mi_n, \aa_n, \WW + \BB_n, \vi_n)$
    \EndFor
    \State \textbf{return} $\mmu_{0|0}, \ldots, \mmu_{N|N}, \SSi_{0|0}, \ldots, \SSi_{N|N}$
    \EndProcedure
  \end{algorithmic}
\end{algorithm}

\subsection{Model Interrogations}\label{sec:mi}

At the heart of the Kalman ODE solver is the specification of the model interrogation in line~\ref{ln:inter} of Algorithm~\ref{alg:Kalman}.  Several proposals from the literature are summarized in Table~\ref{tab:inter}.
\begin{table}[!htb]
    \caption{Various model interrogation methods.}\label{tab:inter}
    \centering
    \resizebox{\columnwidth}{!}{
    \begin{tabular}{llll}
        \toprule
        Reference & $\aa_n$ & $\BB_n$ & $\vi_n$ \\
        \midrule
        \cite{kersting.hennig16} & $-\E[\ff(\xx_n, t_n) \mid \mi_{1:n-1} = \bz]$ & $\bz$ & $\var(\ff(\xx_n, t_n) \mid \mi_{1:n-1} = \bz)$ \\
        \cite{chkrebtii.etal16} &
        $-\ff(\xx^\star_n, t_n): \xx^\star_n \sim \N(\mmu_{n|n-1}, \SSi_{n|n-1})$ & $\bz$
        & $\WW \SSi_{n|n-1}\WW'$ \\
        \cite{schober.etal19} & $-\ff(\mmu_{n|n-1}, t_n)$ & $\bz$ &$\bz$ \\
        \cite{tronarp.etal18} & $-\ff(\mmu_{n|n-1}, t_n) + \JJ_f(\mmu_{n|n-1}, t_n)\mmu_{n|n-1}$ & $-\JJ_f(\mmu_{n|n-1}, t_n)$ & $\bz$ \\
        \cite{kramer21} & $-\ff(\mmu_{n|n-1}, t_n) + \JJ^*_f(\mmu_{n|n-1}, t_n)\mmu_{n|n-1}$ & $-\JJ^*_f(\mmu_{n|n-1}, t_n)$ & $\bz$ \\
        \bottomrule
    \end{tabular}}
\end{table}
The method of~\cite{kersting.hennig16} is shown to be optimal in a class of Gaussian filtering approximations~\citep{tronarp.etal18} and is therefore the most accurate.  However, it requires evaluation of the intractable integrals by numerical methods, e.g., Gaussian quadrature~\citep{kersting.hennig16}.  The method of~\cite{chkrebtii.etal16} can be viewed as approximating these integrals by Monte Carlo~\citep{schober.etal19}.  
The method of~\cite{schober.etal19} is the simplest and fastest.
Moreover, the posterior $p_\L(\xx_{1:N} \mid \mi_{1:N} = \bz)$ is a deterministic function of the inputs to the ODE solver, unlike the Monte Carlo method of~\cite{chkrebtii.etal16}.  However, it has been noted in~\cite{chkrebtii.etal16} that noise-free model interrogation with $\vi_n = \bz$ causes the solver's performance to deteriorate, (which we have confirmed in numerical simulations not presented here). The method of~\cite{tronarp.etal18} uses a first order Taylor expansion instead of the zeroth order of~\cite{schober.etal19} -- with $\JJ_f(\xx,t) = \pdv{\xx} \ff(\xx,t) $ being the Jacobian -- which has been shown to have better numerical stability~\citep{tronarp.etal18}. 
The method of~\cite{kramer21} is a modified version of~\cite{tronarp.etal18}, 
in which the Jacobian is replaced with a modified version $\JJ^*_f(\xx, t)$, which is best explained in the context of solving multivariate ODEs in Section~\ref{sec:block}.

\subsection{Gaussian Markov Process Prior} \label{sec:prior}

A simple and effective prior proposed by~\cite{schober.etal19} assumes that $x(t) = x^{(0)}(t)$ is $q-1$-times integrated Brownian motion (IBM), such that
\begin{equation}\label{eq:ibm}
    x^{(q)}(t) = \sigma B(t).
\end{equation}
This results in a $q$-dimensional continuous Gaussian Markov process
\begin{equation}
  \xx(t) = \big(x^{(0)}(t), x^{(1)}(t), \ldots, x^{(q-1)}(t)\big)
\end{equation}
with matrices $\QQ$ and $\RR$ in~\eqref{eq:mi} given by
\begin{equation}\label{eq:ibmprm}
    Q_{ij} = \mathfrak{1}_{i\leq j}\frac{(\dt)^{j-i}}{(j-i)!}, \qquad R_{ij} = \sigma^2\frac{(\dt)^{2q-1-i-j}}{(2q-1-i-j)(q-1-i)!(q-1-j)!}.
\end{equation}
Here we have tacitly assumed that $q = p$, where $p-1$ is the number of derivatives of $x(t)$ appearing in the ODE-IVP~\eqref{eq:hode}.  However, it is often advantageous in practice to set $q > p$ to increase the smoothness of $x(t)$.  This can be done by padding $\vv$ and $\WW$ in~\eqref{eq:hode} with $q-p$ zeros and $q-p$ columns of zeros, respectively.  For $\vv$, a different method of padding is to work out the values of $x^{(k)}(a)$ for $p \le k < q$ by taking derivatives of the ODE in~\eqref{eq:hode}. 

\rodeo supports arbitrary Gaussian Markov process prior through the specification of $\QQ$ and $\RR$.  Alternatives to the IBM process considered in the probabilistic ODE solver literature include the $q$-times integrated Ornstein–Uhlenbeck process and the Mat\'ern process of smoothness $q$~\citep[e.g.,][]{Tronarp.etal21, kersting20}.   However, the $\QQ$ and $\RR$ matrices for these priors are more difficult to compute, and the IBM process prior performed very well in all the numerical examples in  Section~\ref{sec:examples}.
\subsection{Multiple Variables and Blocking} \label{sec:block}
For a multivariable function $\xx(t) = \big(x_1(t), \ldots, x_d(t)\big)$, let $\XX_k(t) = (x^{(0)}_k(t),\ldots, x^{(p_k-1)}_k(t))$ denote $x_k(t) = x_k^{(0)}(t)$ and its first $p_k-1$ derivatives.  An ODE-IVP for $\xx(t)$ can then be written as  
\begin{equation}\label{eq:mode}
    \WW \XX(t) = \ff(\XX(t), t), \qquad \XX(\tmin) = \vv, \qquad t \in [\tmin, \tmax],
\end{equation}
where
\begin{equation}\label{eq:block}
  \XX(t) = \big(\XX_1(t), \ldots, \XX_d(t)\big).
\end{equation}
The Kalman ODE solver described in Algorithm~\ref{alg:Kalman} can be applied almost exactly for multiple ODE variables as for the univariate case.  It is easy to see that Algorithm~\ref{alg:Kalman} scales linearly in the number of evaluation points $N$.  However, it scales as $\bO(d^3)$ in the number of ODE variables $d$, which is a significant drawback for large ODE systems.

Suppose that it is possible to rearrange the rows of $\WW$ (and corresponding outputs of $f(\XX(t), t)$) such that
\begin{equation}
  \WW =
  \begin{bmatrix}
    \WW^{(1)}_{r_1\times p_1} & & \\
                              & \ddots & \\
                              & & \WW^{(d)}_{r_d\times p_d}
  \end{bmatrix}
\end{equation}
is block diagonal.  Furthermore, suppose that the outputs of the interrogation method are also blocked by system variable, i.e.,
\begin{equation}\label{eq:blockinter}
  \aa_n =
  \begin{bmatrix}
    \aa^{(1)}_{p_1\times 1} \\ \vdots \\ \aa^{(d)}_{p_d\times 1}\big)
  \end{bmatrix}, \qquad
  \BB_n =
  \begin{bmatrix}
    \BB^{(1)}_{p_1\times p_1} & & \\
                              & \ddots & \\
                              & & \BB^{(d)}_{p_d\times p_d}
  \end{bmatrix}, \qquad
  \vi_n =
  \begin{bmatrix}
    \vi^{(1)}_{p_1\times p_1} & & \\
                              & \ddots & \\
                              & & \vi^{(d)}_{p_d\times p_d})
  \end{bmatrix}.
\end{equation}
Then by setting the Gaussian Markov prior on $\XX(t)$ to consist of independent priors for each $\XX_k(t)$, $k = 1,\ldots, d$, i.e., such that
\begin{equation}
  \QQ =
  \begin{bmatrix}
    \QQ^{(1)}_{p_1\times p_1} & & \\
                              & \ddots & \\
                              & & \QQ^{(d)}_{p_d \times p_d}
  \end{bmatrix}, \qquad
  \RR =
  \begin{bmatrix}
    \RR^{(1)}_{p_1\times p_1} & & \\
                              & \ddots & \\
                              & & \RR^{(d)}_{p_d \times p_d}
  \end{bmatrix},
\end{equation}
each of the steps in Algorithm~\ref{alg:Kalman} can be performed variable-wise, thus reducing the computational complexity from $\bO(d^3)$ to $\bO(d)$~\citep{kramer21}.  This acceleration technique, together with the smoothing step to approximate $\XX_{0:N}$ by $\mmu_{0:N|N}$, gives \rodeo a low memory footprint particularly advantageous for backpropagation.  That is, while solvers with high memory cost must approximate gradients via the adjoint method~\citep[e.g.,][]{chen.etal18, kidger21}, \rodeo can simply backpropagate through the solver steps which is typically much faster~\citep{kidger21}.

The block interrogation condition~\eqref{eq:blockinter} is satisfied by the interrogation methods of~\cite{chkrebtii.etal16} and~\cite{schober.etal19} in Table~\ref{tab:inter}, but notably not by the method of~\cite{tronarp.etal18}. The issue lies in the Jacobian $\JJ_f(\xx, t)$, which may not necessarily be block diagonal. The method of~\cite{kramer21} addresses this by keeping only the block diagonal components of $\JJ_f(\xx, t)$ to form $\JJ^*_f(\xx, t)$, and setting the remaining components to zero.  There is extensive evidence in our experiments and those of~\cite{kramer21} suggesting there is minimal loss in accuracy compared to using the full Jacobian $\JJ_f(\xx, t)$ while greatly reducing computational complexity.

The Kalman ODE solver for multiple variables with blocking is extremely similar to Algorithm~\ref{alg:Kalman}: one simply replaces $\xx_n$ by $\XX_n$ and performs the Kalman recursions blockwise.  Defining $\vv = (\vv^{(1)}, \ldots, \vv^{(d)})$, $\mmu_{m|n} = (\mmu_{m|n}^{(1)}, \ldots, \mmu_{m|n}^{(d)})$, $\SSi_{m|n} = \diag(\SSi_{m|n}^{(1)}, \ldots, \SSi_{m|n}^{(d)})$, etc., 
for completeness, the exact steps of the \rodeo solver are presented in Algorithm~\ref{alg:KalmanBlock}.
\begin{algorithm}[!htb]
    \caption{The \rodeo probabilistic ODE solver.  \fct{rodeo\_ode\_sim} is used to obtain a draw from $p_\L(\XX_{1:N} \mid \ZZ_{1:N} = \bz)$, whereas \fct{rodeo\_ode\_mv} is used to obtain $\mmu_{0:N|N}$ and $\SSi_{0:N|N}$, where $\mu_{n|N} = E_\L[\XX_{n} \mid \ZZ_{1:N}=\bz]$ and $\SSi_{n|N} = \var_\L(\XX_n \mid \ZZ_{1:N}=\bz)$.  Both use the same forward pass through the model interrogations in \fct{rodeo\_forward}.  The \code{block} versions of the Kalman algorithms are elementary extensions of the non-block versions.  Thus, only the pseudocode for \fct{kalman\_block\_sample} is provided.}\label{alg:KalmanBlock}
    \begin{algorithmic}[1]
      \Procedure{\code{rodeo\_ode\_sim}}{$\WW, \ff(\XX, t), \vv, \QQ, \RR$}
      \State $(\mmu_{0|0}, \SSi_{0|0}), \ldots, (\mmu_{N|N}, \SSi_{N|N}) \gets \mcode{rodeo\_forward}(\WW, \ff(\XX, t), \vv, \QQ, \RR)$
      \For{$k=1:d$}
      \State $\XX_N^{(k)} \sim \N(\mmu_{N|N}^{(k)}, \SSi_{N|N}^{(k)})$ \label{ln:prng}
      \EndFor
      \State $\XX_N \gets (\XX_N^{(1)}, \ldots, \XX_N^{(d)})$
        \For{$n=N-1:0$}
        \State $\XX_n \gets \kbsample(\XX_{n+1}, \mmu_{n|n}, \SSi_{n|n}, \mmu_{n+1|n}, \SSi_{n+1|n}, \QQ)$
        \EndFor
        \State \textbf{return} $\XX_{0:N}$  
        \EndProcedure
        \State
      \Procedure{\code{rodeo\_ode\_mv}}{$\WW, \ff(\XX, t), \vv, \QQ, \RR$}
      \State $(\mmu_{0|0}, \SSi_{0|0}), \ldots, (\mmu_{N|N}, \SSi_{N|N}) \gets \mcode{rodeo\_forward}(\WW, \ff(\XX, t), \vv, \QQ, \RR)$
        \For{$n=N-1:0$}
        \State $\mmu_{n|N}, \SSi_{n|N} \gets \kbsmooth(\mmu_{n+1|N}, \SSi_{n+1|N}, \mmu_{n|n}, \SSi_{n|n}, \mmu_{n+1|n}, \SSi_{n+1|n}, \QQ)$
        \EndFor
        \State \textbf{return} $\mmu_{0:N|N}, \SSi_{0:N|N}$  
        \EndProcedure
        \State
        \Procedure{\code{rodeo\_forward}}{$\WW, \ff(\XX, t), \vv, \QQ, \RR$}
        \State $\mmu_{0|0}, \Sigma_{0|0} \gets \vv, \bz$
        \State $\ZZ_{1:N} \gets \bz$
        \For{$n=1:N$}
        \State $\mmu_{n|n-1}, \SSi_{n|n-1} \gets \kbpredict(\mmu_{n-1|n-1}, \SSi_{n-1|n-1}, \bz, \QQ, \RR)$
        \State $\aa_n, \BB_n, \vi_n \gets \binter(\mmu_{n|n-1}, \SSi_{n|n-1}, \WW, \ff(\XX, t))$ 
        \State $\mmu_{n|n}, \SSi_{n|n} \gets \kbupdate(\mmu_{n|n-1}, \SSi_{n|n-1}, \ZZ_n, \aa_n, \WW + \BB_n, \vi_n)$ \label{ln:filter}
        \EndFor
        \State \textbf{return} $(\mmu_{0|0}, \SSi_{0|0}), \ldots, (\mmu_{N|N}, \SSi_{N|N})$
        \EndProcedure
        \State \Procedure{\kbsample}{$\XX_{n+1}, \mmu_{n|n}, \SSi_{n|n}, \mmu_{n+1|n}, \SSi_{n+1|n}, \QQ$}
        \For{$k=1:d$}
        \State $\XX_n^{(k)} \gets \ksample(\XX_{n+1}^{(k)}, \mmu_{n|n}^{(k)}, \SSi_{n|n}^{(k)}, \mmu_{n+1|n}^{(k)}, \SSi_{n+1|n}^{(k)}, \QQ^{(k)})$
        \EndFor
        \State \textbf{return} $\XX_n = (\XX_n^{(1)}, \ldots, \XX_n^{(d)})$        
        \EndProcedure
    \end{algorithmic}
\end{algorithm}

\section{Parameter Inference} \label{sec:parinf}

The parameter-dependent extension of the ODE-IVP~\eqref{eq:mode} 
is of the form
\begin{equation}\label{eq:pode}
    \WW_\tth \XX(t) = \ff_\tth(\XX(t), t), \qquad \XX(\tmin) = \vv_\tth, \qquad t \in [\tmin, \tmax].
\end{equation}
The corresponding nonlinear state-space model is
\begin{equation}\label{eq:pbnf}
    \begin{aligned}
        \XX_{n+1} \mid \XX_n & \sim \N(\QQ_\eet \XX_n, \RR_\eet) \\
        \ZZ_n & \ind \N(\WW_\tth \XX_n - \ff_\tth(\XX_n, t_n), \vi_n),
    \end{aligned}
\end{equation}
and the surrogate model is
\begin{equation}\label{eq:pmi}
    \begin{aligned}
        \XX_{n+1} \mid \XX_n & \sim \N(\QQ_\eet \XX_n, \RR_\eet) \\
        \ZZ_n & \ind \N((\WW_\tth + \BB_n)\XX_n + \aa_n, \vi_n),
    \end{aligned}
\end{equation}
where the Gaussian Markov process parameters $\QQ_\eet$ and $\RR_\eet$ depend on tuning parameters $\eet$.
The learning problem consists of estimating the unknown parameters $\tth$ which determine $\XX(t)$ in~\eqref{eq:pode} from noisy observations $\YY_{0:M} = (\YY_0, \ldots, \YY_M)$, recorded at times $t = t'_0, \ldots, t'_M$ under the measurement model
\begin{equation}\label{eq:meas}
    \YY_i \ind p(\YY_i \mid \XX(t'_i), \pph).
\end{equation}
Parameter inference is conducted via the likelihood function for $\TTh = (\tth, \pph, \eet)$,
\begin{equation}\label{eq:likepar}
  \begin{aligned}
    \Ell(\TTh \mid \YY_{0:M}) & \propto p(\YY_{0:M} \mid \ZZ_{1:N} = \bz, \TTh) \\
                              & = \int p(\YY_{0:M} \mid \XX_{0:N}, \pph) p(\XX_{1:N} \mid \ZZ_{1:N}= \bz, \tth, \eet) \dd{\XX_{1:N}}.
  \end{aligned}
\end{equation}

Unlike other software for probabilistic ODE solvers, \rodeo is primarily concerned with parameter inference.  Several methods for ODE parameter inference within the \rodeo framework are presented in the subsections below.

\subsection{Basic Method} \label{sec:fastapp}
%

%

A basic approximation to the likelihood function~\eqref{eq:likepar} takes the posterior mean $\mmu_{0:N|N}(\tth, \eet) = \E_\L[\XX_{0:N} \mid \ZZ_{1:N} = \bz, \tth, \eet]$ of Algorithm~\ref{alg:KalmanBlock} and simply plugs it into the measurement model~\eqref{eq:meas}, such that
\begin{equation} \label{eq:fastlike}
    \hat \Ell(\TTh \mid \YY_{0:M}) = \prod_{i=0}^M p(\YY_i \mid \XX_{n(i)} = \mmu_{n(i)|N}(\tth, \eet), \pph),
\end{equation}
where in terms of the ODE solver discretization time points $t = t_0, \ldots, t_N$, $N \ge M$, the mapping $n(\cdot)$ is such that $t_{n(i)} = t'_i$.  A similar approach is proposed in~\cite{schober.etal19} but without the backward pass in Algorithm~\ref{alg:KalmanBlock}.  The basic approximation~\eqref{eq:fastlike} is very simple, but does not propagate the uncertainty in the probabilistic ODE solver to the calculation of the likelihood.

\subsection{Fenrir} \label{sec:fenrir}

The \fenrir method~\citep{tronarp22} extends a likelihood approximation developed in~\cite{kersting20a}. It is applicable to Gaussian measurement models of the form
\begin{equation}\label{eq:measnorm}
  \YY_i \ind \N(\DD_i^{(\pph)} \XX_{n(i)}, \OOm_i^{(\pph)}).
\end{equation}
\fenrir begins by using the surrogate model~\eqref{eq:pmi} to estimate $p_\L(\XX_{1:N} \mid \ZZ_{1:N} = \bz, \tth, \eet)$. This results in a Gaussian non-homogeneous Markov model going backwards in time,
\begin{equation}\label{eq:fenrirback}
    \begin{aligned}
        \XX_N & \sim \N(\bb_N, \CC_N) \\
        \XX_n \mid \XX_{n+1} & \sim \N(\AA_n \XX_n + \bb_n, \CC_n), \\
    \end{aligned}
\end{equation}
where the coefficients $\AA_{0:N-1}$, $\bb_{0:N}$, and $\CC_{0:N}$ can be derived using the Kalman filtering and smoothing recursions~\citep{tronarp22}.  In combination with the Gaussian measurement model~\eqref{eq:measnorm}, the integral in the likelihood function~\eqref{eq:likepar} can be computed analytically.

In order to benefit from the speed increase of blocking, the matrices $\DD_i^{(\pph)}$ and $\OOm_i^{(\pph)}$ in~\eqref{eq:measnorm} must be block diagonal with $d$ blocks of size $s \times p$ and $s \times s$ respectively, where $s$ is the number of measurements per variable. This is illustrated in Example~\ref{sec:effblock} and the exact steps are presented in Algorithm~\ref{alg:Fenrir}.



\begin{algorithm}[!htb]
    \caption{The Fenrir probabilistic ODE likelihood approximation.}\label{alg:Fenrir}
    \begin{algorithmic}[1]
      \Procedure{\code{fenrir}}{}$\left(
        \begin{aligned}
          & \WW = \WW_\tth, \quad \ff(\XX, t) = \ff_\tth(\XX, t), \quad \vv = \vv_\tth, \\
          & \QQ = \QQ_\eet, \quad \RR = \RR_\eet, \\
          & \YY_{0:M}, \quad \DD_{0:M} = \DD^{(\pph)}_{0:M}, \quad \OOm_{0:M} = \OOm^{(\pph)}_{0:M}
        \end{aligned}
      \right)$
        \State $\ell \gets 0$ \Comment{Initialization}
        \State $i \gets M-1$ \Comment{Used to map $t_{n(i)}$ to $t'_i$}
        \State $\mmu_{0|0}, \ldots, \mmu_{N|N}, \SSi_{0|0}, \ldots, \SSi_{N|N} \gets \mcode{rodeo\_forward}(\WW, \ff(\XX, t), \vv, \QQ, \RR)$
        \State \Comment{Lines 6-21 compute $\log p(\YY_{0:M} \mid \ZZ_{1:N} = \bz, \TTh)$}
        \State $\AA_N, \bb_N, \CC_N \gets \bz, \mmu_{N|N}, \SSi_{N|N}$
        \State $\mmu_N, \SSi_N \gets \kbforecast(\mmu_{N|N}, \SSi_{N|N}, \bz, \DD_M, \OOm_M)$
        \State $\ell \gets \ell + \nlogpdf(\yy_N; \mmu_N, \SSi_N)$
        \State $\mmu_{N|N}^\prime, \SSi_{N|N}^\prime \gets  \kbupdate(\mmu_{N|N}, \SSi_{N|N}, \YY_M, \bz, \DD_M, \OOm_M)$
        \For{$n=N-1:0$}
        \State $\AA_n, \bb_n, \CC_n \gets \kbcond(\mmu_{n|n}, \SSi_{n|n}, \mmu_{n+1|n}, \SSi_{n+1|n}, \QQ)$
        \State $\mmu_{n|n+1}^\prime, \SSi_{n|n+1}^\prime \gets \kpredict(\mmu_{n+1|n+1}^\prime, \SSi_{n+1|n+1}^\prime, \bb_n, \AA_n, \CC)$
        \If{$t_n = t_{n(i)}$}
        \State $\mmu_n, \SSi_n \gets \kbforecast(\mmu_{n|n+1}^\prime, \SSi_{n|n+1}^\prime, \bz, \DD_i, \OOm_i)$
        \State $\ell \gets \ell + \nlogpdf(\YY_i; \mmu_n, \SSi_n)$ 
        \State $\mmu_{n|n}^\prime, \SSi_{n|n}^\prime \gets \kbupdate(\mmu_{n|n+1}^\prime, \SSi_{n|n+1}^\prime, \yy_n, \bz, \DD_i, \OOm_i)$ 
        \State $i \gets i - 1$
        \Else 
        \State $\mmu_{n|n}^\prime, \SSi_{n|n}^\prime \gets \mmu_{n|n+1}^\prime, \SSi_{n|n+1}^\prime$
        \EndIf
        \EndFor
        \State \textbf{return} $\ell$ 
        \EndProcedure
    \end{algorithmic}
\end{algorithm}

\subsection{DALTON}\label{sec:dalton}

The \fenrir approximation can be applied to any of the model interrogations in Table~\ref{tab:inter}.  However, each of these model interrogation methods is data-free, in the sense that it does not use any of the noisy observations $\YY_{0:M}$ in the ODE solver.  In contrast, the \dalton approximation~\citep{wu.etal24} is data-adaptive in that it uses the $\YY_{0:M}$ to approximate the ODE solution.  \dalton provides two methods for computing~\eqref{eq:likepar} depending on whether the measurement model~\eqref{eq:meas} is Gaussian or non-Gaussian. The former uses the identity
\begin{equation}\label{eq:condp}
    p(\YY_{0:M} \mid \ZZ_{1:N} = \bz, \TTh) = \frac{p(\YY_{0:M}, \ZZ_{1:N} = \bz \mid \TTh)}{p(\ZZ_{1:N} = \bz \mid \TTh)}.
\end{equation}
The denominator $p(\ZZ_{1:N} = \bz \mid \TTh)$ on the right-hand side is estimated using a Kalman filter on the data-free surrogate model~\eqref{eq:pmi}~\citep[e.g.,][and see details in Algorithm~\ref{alg:dalton}]{tronarp.etal18, schober.etal19}. The numerator $p(\YY_{0:M}, \ZZ_{1:N} = \bz \mid \TTh)$ can be computed by augmenting the surrogate model~\eqref{eq:pmi} at the measurement locations $n = n(i)$ via
\begin{equation}\label{eq:daltonlin}
    \begin{aligned}
        \begin{bmatrix}
            \ZZ_{n(i)} \\
            \YY_i
        \end{bmatrix} & \ind 
        \N\left(\left(\begin{bmatrix}
            \WW_\tth \\
            \DD^{(\pph)}_i
        \end{bmatrix} + \begin{bmatrix}
            \BB_{n(i)} \\
            \bz
        \end{bmatrix}\right)\XX_{n(i)} + \begin{bmatrix} \aa_{n(i)} \\ \bz \end{bmatrix}, 
        \begin{bmatrix}
            \VV_{n(i)} & \bz \\
            \bz & \OOm^{(\pph)}_i
        \end{bmatrix} \right),
    \end{aligned}
\end{equation}
where the data-free interrogation coefficients $\aa_{n(i)}$, $\BB_{n(i)}$, and $\VV_{n(i)}$ are computed exactly as in Table~\ref{tab:inter}. Additionally, \dalton can also benefit from blocking by having the same requirements for $\DD_i^{(\pph)}$ and $\OOm_i^{(\pph)}$ as \fenrir. The exact computations are presented in Algorithm~\ref{alg:dalton}.  The non-Gaussian case is somewhat more involved and detailed in Appendix~\ref{sec:daltonng}.

\begin{algorithm}[!htb]
    \caption{DALTON probabilistic ODE likelihood approximation for Gaussian measurements.}\label{alg:dalton}
    \begin{algorithmic}[1]
        \Procedure{\code{dalton}}{}$\left(
        \begin{aligned}
          & \WW = \WW_\tth, \quad \ff(\XX, t) = \ff_\tth(\XX, t), \quad \vv = \vv_\tth, \\
          & \QQ = \QQ_\eet, \quad \RR = \RR_\eet, \\
          & \YY_{0:M}, \quad \DD_{0:M} = \DD^{(\pph)}_{0:M}, \quad \OOm_{0:M} = \OOm^{(\pph)}_{0:M}
        \end{aligned}
      \right)$
        \State $\mmu_{0|0}, \Sigma_{0|0} \gets \vv, \bz$ \Comment{Initialization}
        \State $\ZZ_{1:N} \gets \bz$
        \State $\ell_z, \ell_{yz} \gets 0, 0$
        \State $i \gets 1$ \Comment{Used to map $t_n$ to $t'_i$}
        \State \Comment{Lines 7-12 compute $\log p(\ZZ_{1:N} = \bz \mid \TTh)$}
        \For{$n=1:N$}
        \State $\mmu_{n|n-1}, \SSi_{n|n-1} \gets \kbpredict(\mmu_{n-1|n-1}, \SSi_{n-1|n-1}, \bz, \QQ, \RR)$
        \State $\aa_n, \BB_n, \VV_n \gets \binter(\mmu_{n|n-1}, \SSi_{n|n-1}, \WW, \ff(\XX, t_n))$
        \State $\mmu_n, \SSi_n \gets \kbforecast(\mmu_{n|n-1}, \SSi_{n|n-1}, \aa_n, \WW + \BB_n, \VV_n)$
        \State $\ell_z \gets \ell_z + \nlogpdf(\ZZ_n^{(k)}; \mmu_n^{(k)}, \SSi_n^{(k)})$ 
        \State $\mmu_{n|n}, \SSi_{n|n} \gets \kbupdate(\mmu_{n|n-1}, \SSi_{n|n-1}, \ZZ_n, \aa_n, \WW + \BB_n, \VV_n)$
        \EndFor
        \State \Comment{Lines 14-25 compute $\log p(\YY_{0:M}, \ZZ_{1:N} = \bz \mid \TTh)$}
        \State $\ell_{yz} \gets \nlogpdf(\YY_0; \DD_0\vv, \OOm_0)$
        \For{$n=1:N$}
        \State $\mmu_{n|n-1}, \SSi_{n|n-1} \gets \kbpredict(\mmu_{n-1|n-1}, \SSi_{n-1|n-1}, \bz, \QQ, \RR)$
        \State $\aa_n, \BB_n, \VV_n \gets \inter(\mmu_{n|n-1}, \SSi_{n|n-1}, \WW, \ff(\XX, t_n))$
        \If{$t_n = t_{n(i)}$}
        \For{$k=1:d$}
        \State $\ZZ_n^{(k)} \gets  \begin{bmatrix}
            \ZZ_n^{(k)} \\
            \YY_i^{(k)}
        \end{bmatrix}, \qquad
        \WW^{(k)} \gets \begin{bmatrix} 
            \WW^{(k)} \\
            \DD_i^{(k)}
        \end{bmatrix}, \qquad
        \BB_n^{(k)} \gets \begin{bmatrix}
            \BB_n^{(k)} \\
            \bz
        \end{bmatrix}$
        \State $\aa_n^{(k)} \gets \begin{bmatrix}
            \aa_n^{(k)} \\
            \bz
        \end{bmatrix}, \qquad
        \VV_n^{(k)} \gets \begin{bmatrix}
            \VV_n^{(k)} & \bz \\
            \bz & \OOm_i^{(k)}
        \end{bmatrix}$ 
        \State $i \gets i + 1$
        \EndFor
        \EndIf
        \State $\mmu_n, \SSi_n \gets \kbforecast(\mmu_{n|n-1}, \SSi_{n|n-1}, \aa_n, \WW + \BB_n, \VV_n)$    
        \State $\ell_{yz} \gets \ell_{yz} + \nlogpdf(\ZZ_n; \mmu_n, \SSi_n)$
        \State $\mmu_{n|n}, \SSi_{n|n} \gets \kbupdate(\mmu_{n|n-1}, \SSi_{n|n-1}, \ZZ_n, \aa_n, \WW + \BB_n, \VV_n)$
        \EndFor
        \State
        \State \textbf{return} $\ell_{yz} - \ell_z$ \Comment{Estimate of $\log p(\YY_{0:M} \mid \ZZ_{1:N} = \bz, \TTh)$}
        \EndProcedure
    \end{algorithmic}
\end{algorithm}

\subsection{Marginal MCMC} \label{sec:chkmcmc}

In principle, the \fenrir and \dalton algorithms can be applied with any of the model interrogations in Table~\ref{tab:inter}.  
However, the interrogation method of~\cite{chkrebtii.etal16} produces the $\aa_n$ stochastically, such that a direct application of \fenrir would not include the additional uncertainty resulting from this step.  To address this,~\cite{chkrebtii.etal16} adopt a Bayesian approach, whereby for a given parameter prior $\pi(\TTh)$ and arbitrary measurement model~\eqref{eq:meas}, they present a marginal MCMC algorithm for
\begin{equation}
  \begin{aligned}
    \hat{p}(\TTh, \XX_{1:N} \mid \YY_{0:M}, \ZZ_{1:N} = \bz) & \propto \pi(\TTh) \times p(\YY_{0:M} \mid \XX_{0:N}, \pph) \\
                                                             & \times \int \hat{p}_\L(\XX_{1:N}, \ipar_{1:N} \mid \ZZ_{1:N} = \bz, \tth, \eet) \dd{\ipar_{1:N}},
    \end{aligned}
\end{equation}
where $\hat{p}_\L(\XX_{1:N}, \ipar_{1:N} \mid \ZZ_{1:N} = \bz, \tth, \eet)$ corresponds to the surrogate model~\eqref{eq:pmi} with possibly stochastic model interrogation terms $\ipar_{1:N} = (\aa_{1:N}, \BB_{1:N}, \VV_{1:N})$.  The relevant Metropolis-Hastings update is described in Algorithm~\ref{alg:Chkrebtii}.  The term ``marginal'' refers to the fact that $\hat{p}_\L(\XX_{1:N}, \ipar_{1:N} \mid \ZZ_{1:N} = \bz, \tth, \eet)$ appears both in the target posterior density and the Markov transition kernel, thus dropping out of the Metropolis-Hastings acceptance rate completely.  For the Gaussian measurement model~\eqref{eq:measnorm} and deterministic interrogation terms $\ipar_{0:N}$, the marginal MCMC algorithm targets the posterior distribution of the \fenrir likelihood combined with the given prior $\pi(\TTh)$.
\begin{algorithm}[!htb]
    \caption{Marginal MCMC algorithm of~\cite{chkrebtii.etal16}.}\label{alg:Chkrebtii}
    \begin{algorithmic}[1]
      \Procedure{\code{marginal\_mcmc\_step}}{$\TTh^\curr, \XX_{0:N}^\curr, \YY_{0:M}$}
        \State $\TTh^\prop \sim q(\TTh \mid \TTh^\curr)$
        \State $\XX_{0:N}^\prop \gets \mcode{rodeo\_ode\_sim}(\WW_{\tth^\prop}, f_{\tth^\prop}(\XX, t), \vv_{\tth^\prop}, \QQ_{\eet^\prop}, \RR_{\eet^\prop})$ \label{ln:rwkernel}
        \State $\displaystyle \rho \gets \frac{p(\YY_{0:M} \mid \XX_{n(0:M)}^\prop, \eet^\prop) \times \pi(\TTh^\prop) / q(\TTh^\prop \mid \TTh^\curr)}{ p(\YY_{0:M} \mid \XX_{n(0:M)}^\curr, \eet^\curr) \times \pi(\TTh^\curr) / q(\TTh^\curr \mid \TTh^\prop)}$
        \State $v \sim \operatorname{Uniform}(0,1)$
        \If{$v < \rho$}
        \State $\TTh^\curr, \XX_{0:N}^\curr \gets \TTh^\prop, \XX_{0:N}^\prop$
        \EndIf
        \State \textbf{return} $\TTh^\curr, \XX_{0:N}^\curr$ 
        \EndProcedure
    \end{algorithmic}
\end{algorithm}

\subsection{MAGI} \label{sec:magi}

\magi is a method for performing Bayesian inference with the exact posterior
\begin{equation}
    p(\TTh \mid \ZZ_{1:N} = \bz, \YY_{0:M}),
\end{equation}
corresponding to the nonlinear state-space model in~\eqref{eq:pbnf} with $\VV_n = \bz$. This approach differs from the original \magi algorithm proposed by~\cite{yang.etal21} in two key respects. First, the original algorithm is limited to first-order ODEs, whereas our approach generalizes to ODEs of arbitrary order. Second, the original method only assumes a Gaussian process prior on the solution process $\XX_{0:N}$, but not necessarily a Markov one, which substantially increases computational burden. In contrast, by additionally assuming that the prior on $\XX_{0:N}$ has the Markov property, we demonstrate how the \magi posterior can be efficiently and scalably computed within the \rodeo framework.

The \magi approach provides a closed form for the joint posterior
\begin{equation}\label{eq:magijoint}
\begin{aligned}
    p(\TTh, \XX_{1:N} \mid \ZZ_{1:N} = \bz, \YY_{0:M}) &\propto p(\TTh, \XX_{1:N}, \ZZ_{1:N} = \bz, \YY_{0:M}) \\
    &=\pi(\TTh) \times p(\XX_{1:N}, \ZZ_{1:N} = \bz \mid \TTh) \times p(\YY_{0:M} \mid \XX_{0:N}, \pph).
\end{aligned}
\end{equation}
The first and third term on the right-hand side of~\eqref{eq:magijoint} are straightforward to compute.  However, notice that the second term $p(\XX_{1:N}, \ZZ_{1:N} \mid \TTh)$ is a degenerate distribution, since $\ZZ_{1:N} = \WW_{\tth} \XX_{1:N} - \ff_{\tth}(\XX_{1:N}, t_{1:N})$ is uniquely determined by $\XX_{1:N}$.  An significant contribution of~\cite{yang.etal21} is to derive the density of this degenerate distribution.  We present this derivation is a slightly more general setting below.

Assume that $\WW_{\tth} = \WW$ and that $\XX_{0:N}$ can be decomposed as $\XX_{0:N} = (\UU_{0:N}, \WW_{0:N})$, where $\WW_{0:N} = \WW \XX_{0:N}$ is a proper subset of $\XX_{0:N}$ and $\UU_{0:N}$ represents the remaining components of $\XX_{0:N}$, such that for $\ZZ_n = \bz$ we have $\WW_n = \ff_\tth(\UU_n, t_n)$.  For example, consider the parameter estimation problem corresponding to a simple univariate first-order ODE
\begin{equation}\label{eq:exode}
  \begin{aligned}
    x^{(1)}(t) & = \ff_\tth(x^{(0)}(t), t), \\
    \YY_i & \ind p(\YY_i \mid x^{(0)}(t'_i), \pph).
  \end{aligned}
\end{equation}
Then assuming a twice-differetiable IBM prior on the solution process, we have
\begin{equation}
  \XX_n = \begin{bmatrix} x^{(0)}_n \\ x^{(1)}_n \\ x^{(2)}_n \end{bmatrix},
\end{equation}
and the decomposition $\XX_n= (\WW_n, \UU_n)$ is given by 
\begin{equation}
  \WW_n = \begin{bmatrix}x^{(1)}_n \end{bmatrix}, \qquad \UU_n = \begin{bmatrix} x^{(0)}_n \\ x^{(2)}_n \end{bmatrix}.
\end{equation}

Rather than attempting to directly derive the density of the degenerate distribution $p(\XX_{1:N}, \ZZ_{1:N} \mid \TTh)$, note that $(\XX_{0:N}, \ZZ_{1:N})$ can be completely characterized by $(\UU_{0:N}, \ZZ_{1:N})$.
Therefore, we instead consider the nondegenerate joint 
distribution
\begin{equation} 
    p(\UU_{1:N}, \ZZ_{1:N} \mid \TTh),
\end{equation}
which can be obtained by change-of-variables from $\XX_{0:N} = (\UU_{0:N}, \WW_{0:N})$. 
To perform this transformation, we compute the determinant of the Jacobian for the mapping $(\UU_{0:N}, \WW_{0:N}) \to (\UU_{0:N}, \ZZ_{1:N})$, given by
\begin{equation}
\begin{vmatrix}
  \displaystyle \dv{\UU_{0:N}}{\UU_{0:N}} & \displaystyle \dv{\ZZ_{1:N}}{\UU_{0:N}} \\
\displaystyle \dv{\UU_{0:N}}{\WW_{0:N}} & \displaystyle \dv{\ZZ_{1:N}}{\WW_{0:N}}
\end{vmatrix} = 
\begin{vmatrix} 
\Id & \displaystyle \dv{\ZZ_{1:N}}{\UU_{0:N}} \\
\bz & \Id
\end{vmatrix} = 1,
\end{equation}
such that
\begin{equation}\label{eq:magimarkov}
  \begin{aligned}
    p(\UU_{1:N}, \ZZ_{1:N} \mid \TTh)
    & = p(\XX_{1:N} \mid \TTh) \\
    & = \prod_{n=1}^N \varphi(\XX(\UU_n, \ZZ_n) - \QQ_\eet \XX(\UU_{n-1}, \ZZ_{n-1}) \mid \RR_\eet),
  \end{aligned}
\end{equation}
where $\XX(\UU_n, \ZZ_n) = (\UU_n, \ZZ_n + \ff_\tth(\UU_n, t_n))$ and $\varphi(\zz \mid \SSi)$ is the density of a multivariate normal with mean zero and variance $\SSi$.  In other words, $p(\UU_{1:N}, \ZZ_{1:N} \mid \TTh)$ is the density of the Gaussian Markov prior in~\eqref{eq:pbnf} evaluated at $\XX_{0:N} = \XX(\UU_{0:N}, \ZZ_{1:N})$. 
Thus, rewriting~\eqref{eq:magijoint} in terms of the transformed variables $(\UU_{0:N}, \ZZ_{1:N})$, we have
\begin{equation}\label{eq:magipost}
  \begin{aligned}
    p(\TTh, \UU_{1:N} \mid \ZZ_{1:N} = \bz, \YY_{0:M}) & \propto p(\TTh, \UU_{1:N}, \ZZ_{1:N} = \bz, \YY_{0:M}) \\
    & = \pi(\TTh) \times p(\UU_{1:N}, \ZZ_{1:N} = \bz \mid \TTh) \times p(\YY_{0:M} \mid \XX_{0:N}, \pph),
  \end{aligned}
\end{equation}
with the second term on the right-hand side given by~\eqref{eq:magimarkov}.


Sampling from the \magi posterior in~\eqref{eq:magipost} is done via MCMC, which as a byproduct produces samples from $p(\TTh \mid \ZZ_{1:N} = \bz, \YY_{0:M})$. However, it is often possible to significantly reduce the number of variables in the MCMC sampler. That is, consider the smallest subset $\tilde \UU_{0:N} \subseteq \UU_{0:N}$ such that, for $\tilde \XX_{0:N} = (\tilde \UU_{0:N}, \WW_{0:N})$, we have
\begin{equation}
  \begin{aligned}
    \ff_\tth(\tilde \UU_{0:N}, t_{0:N}) & = \ff_\tth(\UU_{0:N}, t_{0:N}), \\
    p(\YY_{0:M} \mid \tilde \XX_{0:N}, \pph) & = p(\YY_{0:M} \mid \XX_{0:N}, \pph).
  \end{aligned}
\end{equation}
Continuing with the example in~\eqref{eq:exode}, we have 
\begin{equation}
\tilde \UU_n = \begin{bmatrix} x^{(0)}_n \end{bmatrix}, \qquad
\tilde\XX_n = \begin{bmatrix} 
        x^{(0)}_n \\ x^{(1)}_n
        \end{bmatrix}.
\end{equation}
The \magi posterior in terms of the smallest subset $\tilde{\UU}_{0:N}$ is
\begin{equation}\label{eq:magimodified}
  \begin{aligned}
    p(\TTh, \tilde \UU_{1:N} \mid \YY_{0:M}, \ZZ_{1:N} = \bz) & \propto \pi(\TTh) \times p(\tilde \XX_{1:N} \mid \TTh) \times p(\YY_{0:M} \mid \tilde \XX_{0:N}, \pph), \\
    & = \pi(\TTh) \times p(\tilde \XX(\tilde \UU_{1:N}) \mid \eet) \times p(\YY_{0:M} \mid \tilde \XX(\tilde \UU_{0:N}), \pph),
    \end{aligned}
\end{equation}
where $\tilde \XX(\tilde \UU_{n}) = (\tilde{\UU}_n, \ff_{\tth}(\tilde{\UU}_n, t_n))$.  In other words, the second term in~\eqref{eq:magimodified} is the density induced by the Gaussian Markov process prior in~\eqref{eq:pbnf} on $\tilde \XX_{0:N} = (\tilde \UU_{0:N}, \WW_{0:N})$, evaluated at $\tilde \XX(\tilde \UU_{0:N})$.

The $\tilde \XX_{1:N}$ in~\eqref{eq:magimodified} are typically not Markov. However, since $\tilde \XX_{1:N}$ is a subset of the Gaussian Markov random variables $\XX_{1:N}$, we can use Kalman recursions to efficiently compute the joint density $p(\tilde \XX_{1:N} \mid \eet)$ with linear time complexity $\bO(N)$.  Thus, the \magi posteriors~\eqref{eq:magipost} and~\eqref{eq:magimodified} for the full and subset variables $\UU_{1:N}$ and $\tilde \UU_{1:N}$ have the same computational complexity per evaluation, but the latter contains $c \times N$ fewer random variables, where $c$ is a function of the number of system variables $d$ and the number of additional derivatives $q-p$ in the IBM prior~\eqref{eq:ibm}.  This can be a significant advantage for speeding up the convergence of the MCMC sampler on~\eqref{eq:magimodified}.

In practice, the length of the solution process $N$ is often much greater than the number of observations $M$, in which case the modified \magi posterior in~\eqref{eq:magimodified} is dominated by the second term~\citep{yang.etal21}. This occurs because the density of $p(\tilde \XX_{1:N} \mid \eet)$ includes many more terms than the likelihood $p(\YY_{0:M} \mid \tilde \XX_{0:N}, \pph)$. To address this imbalance,~\cite{yang.etal21} suggests tempering the density $p(\tilde \XX_{1:N} \mid \eet)^{1/\beta}$ by a hyperparameter $\beta$, referred to as the prior temperature, to better balance the contributions of the density and the likelihood.  The $\beta$-modulated \magi posterior is thus
\begin{equation}\label{eq:magimodbeta}
  \begin{aligned}
    p_\beta(\TTh, \tilde \UU_{1:N} \mid \ZZ_{1:N}, \YY_{0:M})
    & \propto p_\beta(\TTh, \tilde \UU_{1:N}, \ZZ_{1:N} = \bz, \YY_{0:M}) \\
    & = \pi(\TTh) \times p(\tilde \XX(\tilde \UU_{1:N}) \mid \eet)^{1/\beta} \times p(\YY_{0:M} \mid \tilde \XX(\tilde \UU_{0:N}), \pph).
  \end{aligned}
\end{equation}
The exact steps for computing~\eqref{eq:magimodbeta} are presented in Algorithm~\ref{alg:magi}.
\begin{algorithm}[!htb]
    \caption{\magi algorithm of~\cite{yang.etal21} for computing the $\beta$-modulated posterior~\eqref{eq:magimodbeta} with Gaussian Markov process prior.}\label{alg:magi}
    \begin{algorithmic}[1]
        \Procedure{\code{magi\_logpost}}{$\tilde \UU_{0:N}, \ff(\XX, t) = \ff_\tth(\XX, t), \vv = \vv_\tth, \QQ = \QQ_\eet, \RR = \RR_\eet, \YY_{0:M}, \beta$}
        \State $\mmu_{0|0}, \Sigma_{0|0} \gets \vv, \bz$ \Comment{Lines 2-10 compute $p(\tilde \UU_{0:N}, \ZZ_{1:N} = \bz \mid \TTh)$}
        \State $\ell \gets 0$
        \State $\tilde \XX_{0:N} \gets (\tilde \UU_{0:N}, \ff(\tilde \UU_{0:N}, t_{0:N}))$
        \State $\tilde \WW \gets \begin{bmatrix}\Id_{d\times q} & \bz_{d \times p-q}\end{bmatrix}$ \Comment{The dimensions are $\XX_{d\times p}$ and $\tilde \XX_{d\times q}$.}
        \For{$n=1:N$}
        \State $\mmu_{n|n-1}, \SSi_{n|n-1} \gets \kbpredict(\mmu_{n-1|n-1}, \SSi_{n-1|n-1}, \bz, \QQ, \RR)$
        \State $\mmu_n, \SSi_n \gets \kbforecast(\mmu_{n|n-1}, \SSi_{n|n-1}, \bz, \tilde \WW, \bz)$
        \State $\ell \gets \ell + \nlogpdf(\tilde \XX_n; \mmu_n, \SSi_n)$ 
        \State $\mmu_{n|n}, \SSi_{n|n} \gets \kbupdate(\mmu_{n|n-1}, \SSi_{n|n-1}, \tilde \XX_n, \bz, \tilde \WW, \bz)$
        \EndFor
        \State \Comment{Line 12 computes $\log p_\beta(\TTh, \tilde \UU_{1:N}, \ZZ_{1:N} = \bz, \YY_{0:M})$}
        \State $\ell \gets \log \pi(\TTh) + \frac{1}{\beta} \ell + \log p(\YY_{0:M} \mid \tilde \XX_{0:N}, \pph)$
        \State \textbf{return} $\ell$
        \EndProcedure
    \end{algorithmic}
\end{algorithm}

\section{Implementation} \label{sec:packstruct}

\rodeo is designed to serve two types of users.  The first is statisticians and data scientists wishing to use probabilistic solvers to calibrate ODEs to empirical data.  The second is researchers interested in developing new probabilistic methods of ODE parameter inference within the general \rodeo framework
and optimal tuning of the solution prior~\citep{chkrebtii.etal16}.

To best serve these audiences, \rodeo is written in \python using the \jax library for scientific computing.  \jax is a near drop-in replacement for the widely used \pkg{NumPy} linear algebra library~\citep{numpy}, making use of \rodeo straightforward for most \python users.  In addition to programming convenience, \jax offers state-of-the-art JIT compilation and automatic differentiation capabilities, of which the former offers a substantial performance increase over plain \pkg{NumPy} code, and the latter enables the use of high-performance gradient-based parameter learning algorithms (as shown in Section~\ref{sec:examples}).  In particular, parameter inference using the \rodeo likelihood approximations outlined in Section~\ref{sec:parinf} can be conducted using powerful \jax libraries for optimization (\pkg{JAXopt}~\citep{jaxopt}, \pkg{Optax}~\citep{optax}) and MCMC (\pkg{BlackJAX}~\citep{blackjax}).

There are three main components to the \rodeo library.  The first is a general-purpose Kalman filter and smoother.  The main Kalman algorithms are:
\begin{itemize}
\item \fct{predict}: Given $\mmu_{n-1|n-1}$ and $\SSi_{n-1|n-1}$, calculate $\mmu_{n|n-1}$ and $\SSi_{n|n-1}$.
\item \fct{update}: Given $\mmu_{n|n-1}$ and $\SSi_{n|n-1}$, calculate $\mmu_{n|n}$ and $\SSi_{n|n}$.
\item \fct{smooth}: Given $\mmu_{n+1|N}$ and $\SSi_{n+1|N}$, calculate $\mmu_{n|N}$ and $\SSi_{n|N}$.
\end{itemize}
The standard version of these algorithms (and a few related ones) are detailed in Appendix~\ref{sec:kalmanfun}. However, it has been shown that the so-called square-root variant of these algorithms~\citep{kaminski.etal71, bierman73} can greatly improve numerical accuracy in high-dimensional and/or stiff ODE systems~\citep{kramer20}. Both the standard and square-root of these algorithms are shipped in the \code{kalman} module of the \rodeo libary.

The second primary component of \rodeo is the ODE solver itself.  This consists of two functions, \fct{solve\_mv} and \fct{solve\_sim}, which return the mean and variance $(\mmu_{0:N|N}, \SSi_{0:N|N})$ and a random draw $\XX_{1:N} \sim p(\XX_{1:N} \mid \ZZ_{1:N} = \bz)$, respectively, for the surrogate model~\eqref{eq:pmi}.  The common parameters to these functions are:
\begin{itemize}
\item \code{key}: A \jax pseudo-RNG key.  This is needed for any \jax calculation which requires random samples.
\item \code{ode\_weight}, \code{ode\_init}, \code{ode\_fun}, \code{params}: The ODE-IVP model components $\WW$, $\vv$, $\ff_\tth(\XX, t)$ and $\tth$.  The ODE function has argument signature \code{function(X, t, **params)}, such that the parameters of the model may be passed via arbitrary key-value pairs.  These same key-value pairs must then be supplied to \fct{solve\_sim} and \fct{solve\_mv}.
\item \code{t\_min}, \code{t\_max}, \code{n\_steps}: The time interval $(\tmin, \tmax)$ over which the ODE solution is sought, and the number of solver discretization points $N$.
\item \code{prior\_weight}, \code{prior\_var}: The weight and variance matrices $\QQ$ and $\RR$ of the solution prior as discretized to time intervals of size $\dt = (\tmax - \tmin)/N$.
\item \code{interrogate}: An interrogation function, i.e., a function with arguments \code{key}, $\mmu_{n|n-1}$, $\SSi_{n|n-1}$, $\WW$, $\ff_\tth(\XX, t)$, and $\tth$ which returns the interrogation variables $\aa_n$, $\BB_n$, and $\VV_n$ used in the surrogate model~\eqref{eq:pmi}.  Note that the \code{key} argument is used here to implement stochastic interrogation methods such as that of~\cite{chkrebtii.etal16} in Table~\ref{tab:inter}.
\item \code{kalman\_type}: This argument allows users to select either the standard or square-root implementation of the Kalman algorithms available in \rodeo.
\end{itemize}

The third main component of \rodeo consists of implementations of the Basic, \fenrir, \dalton, marginal MCMC, and \magi inference algorithms described in Section~\ref{sec:parinf}.  These implementations share many of the same arguments as \fct{solve\_sim} and \fct{solve\_mv}.  Additional arguments are:
\begin{itemize}
\item \code{obs\_data}, \code{obs\_times}: The observed data $\YY_{0:M}$ and corresponding observation times $t'_{0:M}$.  For simplicity, these are rounded to the nearest values of the solution grid $t_{0:N}$.
\item \code{obs\_weight}, \code{obs\_var}: For the \fenrir and Gaussian \dalton algorithms, the weight and variance matrices $\DD_{1:M}^{(\pph)}$ and $\OOm_{1:M}^{(\pph)}$ in the multivariate normal observation model~\eqref{eq:measnorm}.
\item \code{obs\_loglik}: For the Basic and Marginal MCMC algorithms, the log of the observation likelihood $p(\YY_{0:M} \mid \XX_{n(0:M)}, \pph)$ in~\eqref{eq:meas}.  This is a function with argument signature \code{function(obs\_data, ode\_data, **params)}, corresponding to $\YY_{0:M}$, $\XX_{n(0:M)}$, and $\pph$, respectively.
\end{itemize}
A few other algorithm-specific arguments will be discussed in the relevant examples in Section~\ref{sec:examples}.

Finally, \rodeo provides a few helper functions.  The two most important ones are \fct{ibm\_init} and \fct{indep\_init} to help set up the prior.  \fct{ibm\_init} creates the weight and variance matrices \code{prior\_weight} and \code{prior\_var} for the solution prior $\XX(t)$, for given time interval $\dt$ and number of required continuous derivatives per ODE variable, $p_1, \ldots, p_d$.

By default, we set all $p_k$, $k = 1,\ldots,d$ equal to their maximum value and return \jax arrays of size $d \times p \times p$, where $p = \max_{1\le k\le d} p_k$.  It is suggested that the ODE-IVP components \code{ode\_weight}, \code{ode\_init}, and \code{ode\_fun} be zero-padded accordingly (see Section~\ref{sec:examples}), and similarly stored in block format, i.e.,  with system variables as the leading dimension of \jax arrays.  This typically offers a performance boost despite padding when there are lots of variables with a similar number of derivatives.


Zero-padding can be disabled in \fct{solve\_mv} and \fct{solve\_sim} by passing the ODE-IVP components as a single block.  To disable zero-padding for the IBM prior, we can compute $\QQ^{(k)}$ and $\RR^{(k)}$ separately for each $k = 1,\dots,d$, then combine them into densely-stored block-diagonal matrices using \fct{indep\_init}. 

\section{Examples} \label{sec:examples}

The following examples illustrate the use of \rodeo for solving and learning the parameters of various ODE-IVPs on $\xx(t) = (x_1(t), \ldots, x_d(t))$. All examples are implemented with variable blocking as described above.  That is, if the ODE-IVP involves the first $p_k-1$ derivatives of $x_k(t)$, then $x_k(t)$ is given an IBM prior with $p = \max_{1\le k \le d} p_k$ continuous derivatives, i.e., $q_k = p+1$, and the ODE-IVP is padded accordingly.  In the context of parameter inference, it will be convenient to introduce a problem-specific parameter transformation $\TTh = \utrans(\tth, \pph, \eet)$ such that the support of $\TTh$ is all of $\mathbb{R}^{\dim\{\tth\} + \dim\{\pph\} + \dim\{\eet\}}$.



\subsection{A Second-Order Univariate ODE}\label{sec:hode}



Consider the second-order univariate ODE-IVP given by
\begin{equation} \label{eq:chkrebtiiode}
  x^{(2)}(t) = \sin(2t) - x(t), \qquad  (x(0), x^{(1)}(0)) = (-1, 0), \qquad t \in [0,10],
\end{equation}
employed by~\cite{chkrebtii.etal16} to illustrate the benefits of using stochastic solvers.
We recast~\eqref{eq:chkrebtiiode} in the form of~\eqref{eq:mode} and go through the inputs required for \rodeo.
Since~\eqref{eq:chkrebtiiode} involves derivatives up to second order, we use a three-times integrated IBM prior on $\XX(t) = (x^{(0)}(t), \ldots, x^{(3)})$.
Since this problem is univariate, we have $d=1$ blocks for which we use an extra dimension to store the blocks.  Thus, using \pkg{NumPy}/\jax array notation, \eqref{eq:chkrebtiiode} is written in the form of~\eqref{eq:mode}  as
\begin{equation}
  \begin{aligned}
    \XX(t) & = \begin{bmatrix}\begin{bmatrix}x^{(0)}(t) & x^{(1)}(t) & x^{(2)}(t) & x^{(3)}(t)\end{bmatrix}\end{bmatrix},  &
        \ff(\XX(t), t) & = \begin{bmatrix}\begin{bmatrix}\sin(2t) - x^{(0)}(t)\end{bmatrix}\end{bmatrix}, \\
        \WW & = \begin{bmatrix}\begin{bmatrix}\begin{bmatrix} 0 & 0 & 1 & 0 \end{bmatrix} \end{bmatrix}\end{bmatrix}, &
        \vv = \XX(0) & = \begin{bmatrix}\begin{bmatrix}-1 & 0 & 1 & 0\end{bmatrix} \end{bmatrix}.
    \end{aligned}
\end{equation}

These variables correspond to \code{ode\_fun}, \code{ode\_weight}, and \code{ode\_init} respectively. The time interval in~\eqref{eq:chkrebtiiode} determines the function arguments \code{t_min = 0} and \code{t_max = 10}. We use \code{n\_steps} to represent the number of solver discretization points. We want \code{n\_steps} to be small so that the run time is fast but large enough to give an accurate approximation. A suitable number satisfying these conditions is \code{n\_steps = 80}. The effect of increasing \code{n\_steps} will be examined shortly. We use the IBM process as our Gaussian prior which require step size $\dt$ and scale parameter $\sigma$. The former is determined by \code{t\_min}, \code{t\_max}, and \code{n\_steps} to be $\dt=10/80$. For the latter, our experiments indicate a remarkable insensitivity to the choice of $\sigma$ in the Gaussian prior over several orders of magnitude, thus for simplicity we set $\sigma=0.1$. We use \code{ibm\_init} to compute \code{prior\_weight} and \code{prior\_var} for the IBM prior. Finally, we set \code{interrogate} to use the~\cite{schober.etal19} and~\cite{chkrebtii.etal16} interrogation method for this example. The code below provides a tutorial for solving this problem. 


\begin{lstlisting}[language=iPython, caption={\rodeo solver for the higher order ODE~\eqref{eq:hode}.}]
# --- 0. Import libraries and module -------------------------------------

import jax
import jax.numpy as jnp
import rodeo
import rodeo.prior
import rodeo.interrogate
import functools  # library for partial

# --- 1. Define the ODE-IVP ----------------------------------------------


def higher_fun(x, t):
    """
    Higher-order ODE of Chkrebtii et al in **rodeo** format.
    Args:
        x: JAX array of shape `(1,4)` corrresponding to
           `X = (x, x^(1), x^(2), x^(3))`.
        t: Scalar time variable.

    Returns:
        JAX array of shape `(1,1)` corresponding to `f(x,t)`.

    """
    return jnp.array([[jnp.sin(2 * t) - x[0, 0]]])


W = jnp.array([[[0.0, 0.0, 1.0, 0.0]]])  # LHS matrix of ODE
x0 = jnp.array([[-1.0, 0.0, 1.0, 0.0]])  # initial value for the IVP

# Time interval on which a solution is sought.
t_min = 0.0
t_max = 10.0

# --- 3.  Define the prior process ---------------------------------------

n_vars = 1  # number of variables in the ODE
n_deriv = 4  # max number of derivatives
sigma = jnp.array([0.1] * n_vars)  # IBM process scale factor


# --- 4.  Evaluate the ODE solution --------------------------------------

n_steps = 80  # number of evaluations steps
dt = (t_max - t_min) / n_steps  # step size

# generate the Kalman parameters corresponding to the prior
prior_Q, prior_R = rodeo.prior.ibm_init(dt=dt, n_deriv=n_deriv, sigma=sigma)

# deterministic ODE solver: posterior mean
Xt, _ = rodeo.solve_mv(
    key=None,  # not needed as interrogate function is deterministic
    # define ode
    ode_fun=higher_fun,
    ode_weight=W,
    ode_init=x0,
    t_min=t_min,
    t_max=t_max,
    # solver parameters
    n_steps=n_steps,
    interrogate=rodeo.interrogate.interrogate_schober,
    prior_weight=prior_Q,
    prior_var=prior_R,
    kalman_type="standard",  # standard Kalman algorithms
)

key = jax.random.PRNGKey(0)  # JAX pseudo-RNG key

Xt = rodeo.solve_sim(
    key=key,
    # define ode
    ode_fun=higher_fun,
    ode_weight=W,
    ode_init=x0,
    t_min=t_min,
    t_max=t_max,
    # solver parameters
    n_steps=n_steps,
    # if using interrogate_chkrebtii, need to fix the value for kalman_type
    # this is done using the functools.partial function
    interrogate=functools.partial(
        rodeo.interrogate.interrogate_chkrebtii, kalman_type="square-root"
    ),
    prior_weight=prior_Q,
    # if using square-root filter, the prior variance needs to be cholesky
    prior_var=jax.vmap(jnp.linalg.cholesky)(prior_R),
    kalman_type="square-root",  # square-root Kalman algorithms
)
\end{lstlisting}
\begin{figure}[H]
    \includegraphics[width=\linewidth]{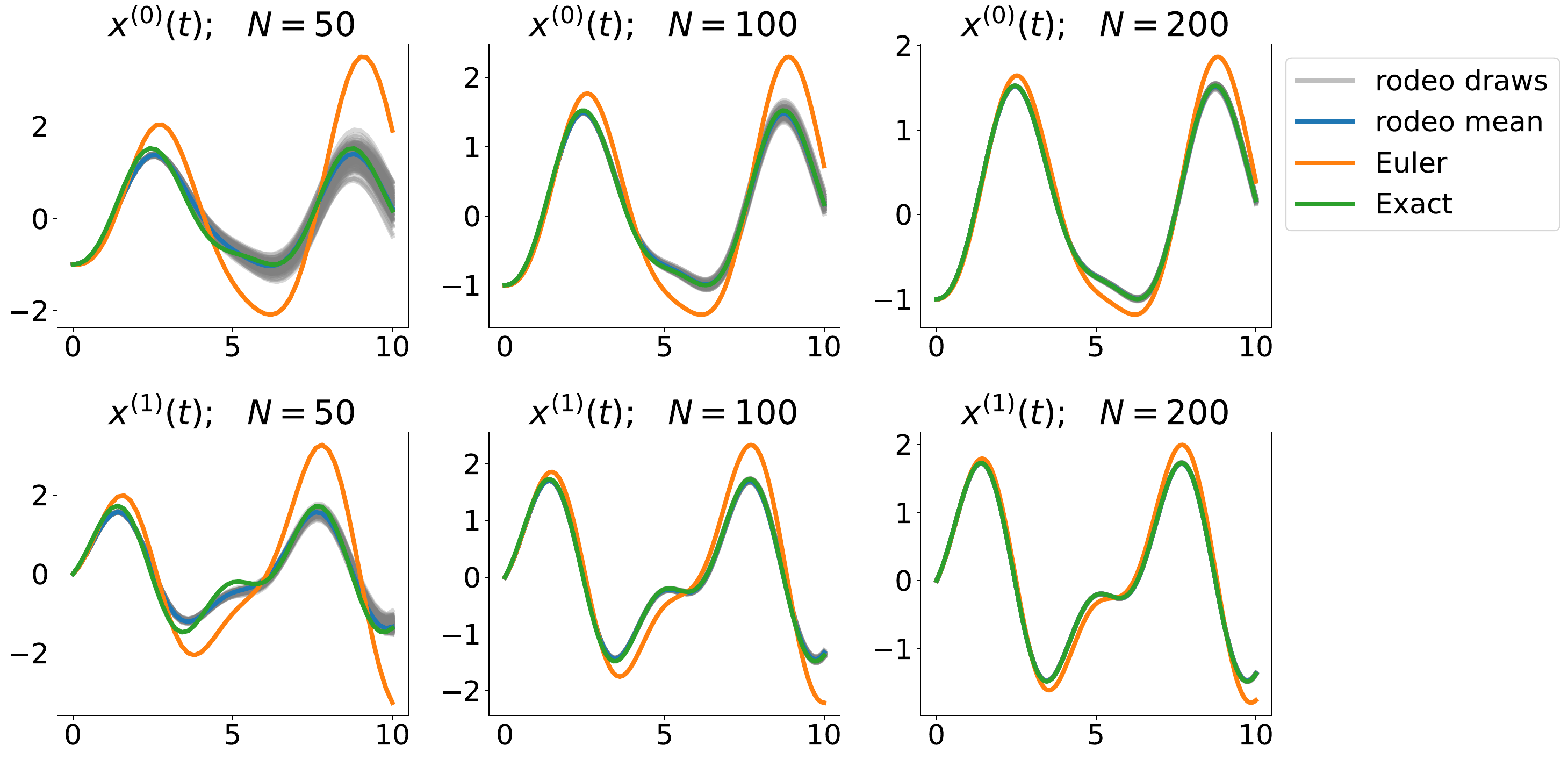}
    \caption{\rodeo, Euler and the exact solution at $N=50, 100, 200$ evaluation points for $t \in [0,10]$. 100 draws and the analytical posterior mean were produced from \rodeo. }
    \label{fig:chkrebtiifig}
\end{figure}

For this problem, a closed-form solution is given by
\begin{equation}
    x^{(0)}(t) = \tfrac 1 3 \big(2\sin(t) - 3\cos(t) - \sin(2t)\big).
\end{equation}
In Figure~\ref{fig:chkrebtiifig}, we use the exact solution to benchmark \rodeo at different numbers of evaluation points $N = 50, 100, 200$.
For each $N$, $100$ independent draws were simulated from \fct{solve\_sim}, and the posterior mean was computed using \fct{solve\_mv}.  Figure~\ref{fig:chkrebtiifig} also displays the ODE solution as approximated by a simple Euler scheme:
\begin{equation}\label{eq:euler}
  \begin{aligned}
    x^{(1)}(t + \dt) & = x^{(1)}(t) + \big(\sin(2t) - x^{(0)}(t)\big) \dt \\
    x^{(0)}(t + \dt) & = x^{(0)}(t) + x^{(1)}(t) \dt.
  \end{aligned}
\end{equation}
As $N$ increases, \rodeo approaches the true solution. \rodeo is more accurate than Euler for any number of evaluation points, most apparently at $N=50$. Figure~\ref{fig:chkrebtiifig} also indicates that \rodeo is able to capture the uncertainty in the solution posterior at low $N$,
where the solver is less accurate.

\subsection{The FitzHugh-Nagumo Model} \label{sec:fitz}

The FitzHugh-Nagumo (FN) model~\citep{Sherwood2013} is a two-state ODE on $\xx(t) = \big(V(t), R(t)\big)$,
in which $V(t)$ describes the evolution of the neuronal membrane voltage and $R(t)$ describes the activation and deactivation of neuronal channels. The FN ODE is given by
\begin{equation} \label{eq:fitz}
  \begin{aligned}
    V^{(1)}(t) &= c\Big(V(t) - \frac{V(t)^3}{3} + R(t)\Big), \\
    R^{(1)}(t) &= -\frac{V(t) - a + bR(t)}{c}.
  \end{aligned}
\end{equation}
In the following examples (Sections~\ref{sec:laplaceapp}-\ref{sec:magiex}), we use various \rodeo inference algorithms to learn the unknown parameters $\tth = (a, b, c, V(0), R(0))$ with $a,b,c > 0$ from data $\YY_{0:M} = (\YY_0, \ldots, \YY_M)$ following the measurement model
\begin{equation}\label{eq:fitznoise}
  \YY_i \ind \N(\xx(t_i), \phi^2 \cdot \Id_{2\times 2}).
\end{equation}
A comparison of these algorithms is provided in Section~\ref{sec:fncomp}.

\begin{figure}[H]
    \centering
    \includegraphics[width=\linewidth]{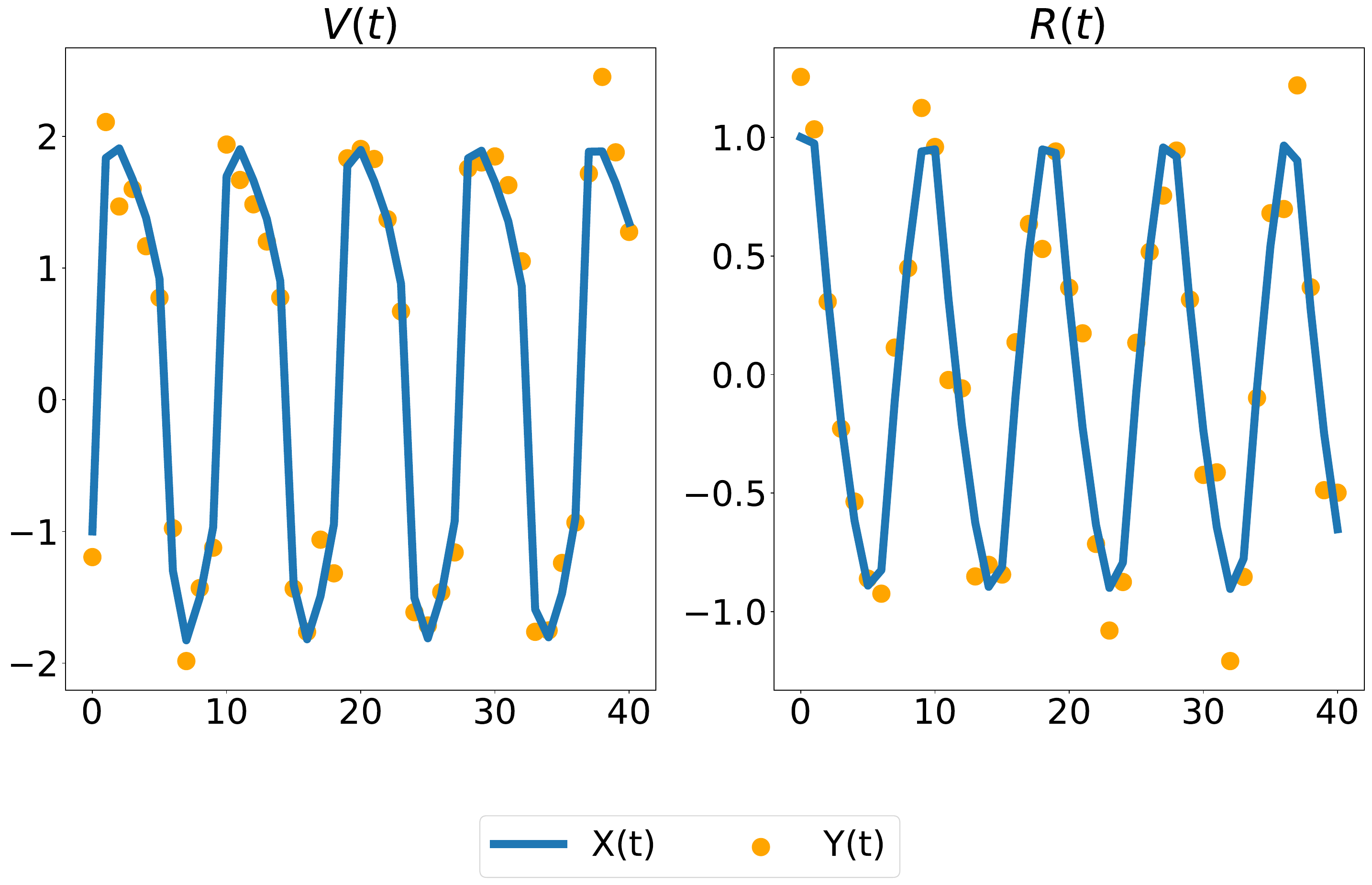}
    \caption{Simulated ODE and noisy observations of the FN model.
    }
    \label{fig:fitzsim}
\end{figure}

We begin by simulating data with $M = 40$, $t_i = i$, $\phi = 0.2$, and true parameters $\tth = (0.2, 0.2, 3, -1, 1)$.  This is accomplished using a very small step size $\dt$ with the interrogation method of~\cite{kramer21} in Table~\ref{tab:inter}, which we found to achieve the highest accuracy as a function of $\dt$.  The simulated ODE trajectory and noisy observations are displayed in Figure~\ref{fig:fitzsim}.

To assist with zero-padding for first-order ODEs using a constant selection matrix $\WW$ in~\eqref{eq:mode} (which does not depend on $\tth$), we provide a helper function \fct{first\_order\_pad}. This function takes three arguments: the ODE function \code{fitz_fun}, the number of variables \code{n_vars}, and the number of derivatives used in the Gaussian Markov process prior \code{n_deriv}. It returns zero-padded versions of both $\WW$ and the ODE function. The padded ODE function can then be used to transform the initial values $\vv_\tth$ into a vector for each variable, stacked into the full state array expected by the \rodeo solver.
\begin{lstlisting}[language=iPython, lastline=93, caption={Noisy observations simulated for the FN model using \rodeo.}]
# --- Import libraries and modules ---------------------------------------

import jax
import jax.numpy as jnp
import rodeo  # ODE solvers
import rodeo.prior  # IBM prior
import rodeo.interrogate  # interrogation methods

# helper function to help with the padding in rodeo format
from rodeo.utils import first_order_pad

jax.config.update("jax_enable_x64", True)

# --- Define the ODE-IVP -------------------------------------------------


def fitz_fun(X, t, theta):
    "FitzHugh-Nagumo ODE in rodeo format."
    a, b, c = theta
    V, R = X[:, 0]
    return jnp.array([[c * (V - V * V * V / 3 + R)], [-1 / c * (V - a + b * R)]])


n_vars = 2
n_deriv = 3

# helper function for standard first-order problems where W is fixed, it returns
# W: LHS matrix of ODE
# fitz_init_pad: Function to help initialize FN model with
# Args: x0, t, **params
# Returns: Function that takes the initial values of each variable
#          and puts them in rodeo format.
W, fitz_init_pad = first_order_pad(fitz_fun, n_vars, n_deriv)
theta = jnp.array([0.2, 0.2, 3])  # ODE parameters
x0 = jnp.array([-1.0, 1.0])  # initial value for the ODE-IVP
X0 = fitz_init_pad(x0, 0, theta=theta)  # initial value in rodeo format


# --- Define the prior process -------------------------------------------

# Time interval on which a solution is sought.
t_min = 0.0
t_max = 40.0

# IBM process scale factor
sigma = jnp.array([0.1] * n_vars)

# --- data simulation ------------------------------------------------------

dt_obs = 1.0  # interobservation time
n_steps_obs = int((t_max - t_min) / dt_obs)
# observation times
obs_times = jnp.linspace(t_min, t_max, num=n_steps_obs + 1)

# number of simulation steps per observation
n_res = 50
n_steps = n_steps_obs * n_res
# simulation times
sim_times = jnp.linspace(t_min, t_max, num=n_steps + 1)

# prior parameters
dt_sim = (t_max - t_min) / n_steps  # grid size for simulation
prior_Q, prior_R = rodeo.prior.ibm_init(dt=dt_sim, n_deriv=n_deriv, sigma=sigma)

# Produce a Pseudo-RNG key
key = jax.random.PRNGKey(0)

# calculate ODE via deterministic output
key, subkey = jax.random.split(key)
Xt, _ = rodeo.solve_mv(
    key=None,
    # define ode
    ode_fun=fitz_fun,
    ode_weight=W,
    ode_init=X0,
    t_min=t_min,
    t_max=t_max,
    theta=theta,  # ODE parameters added here
    # solver parameters
    n_steps=n_steps,
    interrogate=rodeo.interrogate.interrogate_kramer,
    prior_weight=prior_Q,
    prior_var=prior_R,
    kalman_type="standard",  # default choice of Kalman filter
)

# generate observations
noise_sd = 0.2  # Standard deviation in noise model
eps = jax.random.normal(key=subkey, shape=(obs_times.size, 2))
# 0th order derivatives at observed timepoints
obs_ind = jnp.searchsorted(sim_times, obs_times)
x = Xt[obs_ind, :, 0]
Y = x + noise_sd * eps
\end{lstlisting}

\newpage

\subsection{Parameter Inference via Laplace Approximation}\label{sec:laplaceapp}

Parameter estimation with the probabilistic ODE solver employs the likelihood function
\begin{equation}
\Ell(\TTh \mid \YY_{0:M}) \propto p(\YY_{0:M} \mid \ZZ_{1:N} = \bz, \TTh)  
\end{equation}
for an invertible parameter transformation $\TTh = \utrans(\tth, \pph, \eet)$.  In a Bayesian setting, this is combined with a prior distribution $\pi(\TTh)$ 
to obtain the posterior distribution
\begin{equation}\label{eq:postpar}
  p(\TTh \mid \YY_{0:M}) \propto \pi(\TTh) \times \Ell(\TTh \mid \YY_{0:M}).
\end{equation}
For the Basic, \fenrir, and \dalton algorithms described in Section~\ref{sec:parinf}, the high-dimensional latent ODE variables $\XX_{0:N}$ can be approximately integrated out to produce a closed-form likelihood approximation $\hat \Ell(\TTh \mid \YY_{0:M})$ and its corresponding posterior approximation $\hat p(\TTh \mid \YY_{0:M})$.  While this posterior can be readily sampled from using MCMC techniques (as we shall do momentarily), Bayesian parameter estimation can also be achieved by way of a Laplace approximation~\citep[e.g.,][Chapter~4]{gelman.etal13}. Namely, $p(\TTh \mid \YY_{0:M})$ is approximated by a multivariate normal distribution,
\begin{equation}\label{eq:bna}
    \TTh \mid \YY_{0:M} \approx \N(\hat \TTh, \hat \VV),
\end{equation}
where
\begin{equation}\label{eq:mq}
  \begin{aligned}
    \hat \TTh & = \argmax_{\TTh} \log \hat p(\TTh \mid \YY_{0:M}), & \hat \VV & = -\left[\frac{\partial^2}{\partial \TTh \partial \TTh'} \log \hat p(\hat \TTh \mid \YY_{0:M})\right]^{-1}.
  \end{aligned}
\end{equation}
One can then obtain posteriors on the original scale $(\tth, \pph, \eet)$ by first sampling $\TTh^{(1)}, \ldots, \TTh^{(B)}$ from the normal distribution~\eqref{eq:bna}, then computing $(\tth^{(b)}, \pph^{(b)}, \eet^{(b)}) = \utrans^{-1}(\TTh^{(b)})$.

The Laplace approximation is a popular tool for Bayesian machine learning applications~\citep[e.g.,][]{mackay92, gianniotis19}, being typically much faster than MCMC. Our \python implementation of \rodeo uses the \jax library~\citep{jax2018github} for automatic differentiation and JIT-compilation.  This produces very fast implementations of $\log \hat p(\TTh \mid \YY_{0:M})$, along with its gradient and Hessian.

The code below shows how to implement the Laplace approximation -- also called the Bayesian normal approximation -- for estimating the parameter posterior $p(\tth \mid \YY_{0:M})$ of the FN model using the Basic ODE likelihood approximation described in Section~\ref{sec:fastapp}.  We take the measurement model parameter $\phi$ to be known, such that the learning problem is only for the ODE model parameters $\tth = (a, b, c, V(0), R(0))$ and the Gaussian Markov process tuning parameters $\eet = (\sigma_1, \sigma_2)$.  The parameter transformation is
\begin{equation}
  \TTh = \big(\upar_{\tth} = (\log a, \log b, \log c, V(0), R(0)), \upar_{\eet} = (\log \sigma_1, \log \sigma_2)\big).
\end{equation}
The prior $\pi(\TTh)$ consists of independent $\N(0, 10^2)$ distributions on each element of $\upar_{\tth}$ and a flat prior on the tuning parameters $\pi(\eet) \propto 1$.

The posterior mode $(\hat{\upar}_{\tth}, \hat{\upar}_{\eet})$ is obtained using the Newton-CG optimization algorithm~\citep{NoceWrig06} as implemented in the \pkg{JAXopt} library~\citep{jaxopt}. As for the corresponding estimate of the posterior variance of $\upar_{\tth}$, we use only the Hessian of $\log \hat p(\upar_{\tth}, \upar_{\eet} \mid \YY_{0:M})$ in terms of the model parameters $\upar_{\tth}$.  This is due to the insensitivity of the stochastic solver to the values of $\eet$ beyond a rough order of magnitude, which had the effect of unduly inflating the posterior variance estimate of $\upar_{\tth}$ when the Hessian was taken with respect to both $\upar_{\tth}$ and $\upar_{\eet}$.  This approach has been previously advocated in~\cite{tronarp22}.
\begin{remark}
  For the flat prior $\pi(\TTh) \propto 1$, the posterior mean and variance estimates $\hat{\TTh}$ and $\hat{\VV}$ in~\eqref{eq:mq} correspond to the maximum likelihood estimate (MLE) and inverse of the observed Fisher information for the approximate likelihood $\hat \Ell(\TTh \mid \YY_{0:M})$.  The code below can thus be used for Frequentist inference as well, e.g., with the standard errors of $\hat{\TTh}$ given by the square root of the diagonal elements of $\hat{\VV}$. 
\end{remark}

\begin{lstlisting}[language=iPython,firstline=95,lastline=223, caption={Parameter posterior for the FN model using the Laplace approximation with the Basic ODE likelihood approximation.} ]
# --- Import libraries and modules ---------------------------------------

import jax
import jax.numpy as jnp
import jaxopt  # optimization library for finding mode in Laplace approximation
import rodeo.prior  # IBM prior
import rodeo.interrogate  # interrogation methods
import rodeo.inference  # parameter inference methods

# --- parameter inference: basic + laplace -------------------------------


def fitz_logprior(upars):
    "Logprior on unconstrained model parameters."
    n_theta = 5  # number of ODE + IV parameters
    lpi = jax.scipy.stats.norm.logpdf(x=upars[:n_theta], loc=0.0, scale=10.0)
    return jnp.sum(lpi)


def fitz_loglik(obs_data, ode_data, theta):
    """
    Loglikelihood for measurement model.

    Args:
        obs_data (ndarray(n_obs, n_vars)): Observations data.
        ode_data (ndarray(n_obs, n_vars, n_deriv)): ODE solution.
    """
    ll = jax.scipy.stats.norm.logpdf(x=obs_data, loc=ode_data[:, :, 0], scale=noise_sd)
    return jnp.sum(ll)


def fitz_constrain_pars(upars, dt):
    """
    Convert unconstrained optimization parameters into rodeo inputs.

    Args:
        upars : Parameters vector on unconstrainted scale.
        dt : Discretization grid size.

    Returns:
        tuple with elements:
        - theta : ODE parameters.
        - X0 : Initial values in rodeo format.
        - Q, R : Prior matrices.
    """
    theta = jnp.exp(upars[:3])
    x0 = upars[3:5]
    X0 = fitz_init_pad(x0, 0, theta=theta)
    sigma = jnp.exp(upars[5:])
    Q, R = rodeo.prior.ibm_init(dt=dt, n_deriv=n_deriv, sigma=sigma)
    return theta, X0, Q, R


def fitz_laplace(key, neglogpost, n_samples, upars_init):
    """
    Sample from the Laplace approximation to the parameter posterior for the FN model.

    Args:
        key : PRNG key.
        neglogpost: Function specifying the negative log-posterior distribution
                    in terms of the unconstrained parameter vector ``upars``.
        upars_init: Initial value to the optimization algorithm over ``neglogpost()``.
        n_samples : Number of posterior samples to draw.

    Returns:
        JAX array of shape ``(n_samples, 5)`` of posterior samples from ``(theta, x0)``.
    """
    n_theta = 5  # number of ODE + IV parameters
    # find mode of neglogpost()
    solver = jaxopt.ScipyMinimize(
        fun=neglogpost,
        method="Newton-CG",
        jit=True,  # jaxopt jits the neglogpost function for us
    )
    opt_res = solver.run(upars_init)
    upars_mean = opt_res.params
    # variance estimate
    upars_fisher = jax.jacfwd(jax.jacrev(neglogpost))(upars_mean)
    # unconstrained ode+iv parameter variance estimate
    uode_var = jax.scipy.linalg.inv(upars_fisher[:n_theta, :n_theta])
    uode_mean = upars_mean[:n_theta]
    # sample from Laplace approximation
    uode_sample = jax.random.multivariate_normal(
        key=key, mean=uode_mean, cov=uode_var, shape=(n_samples,)
    )
    # convert back to original scale
    ode_sample = uode_sample.at[:, :3].set(jnp.exp(uode_sample[:, :3]))
    return ode_sample


def fitz_neglogpost_basic(upars):
    "Negative logposterior for basic approximation."
    # solve ODE
    theta, X0, prior_Q, prior_R = fitz_constrain_pars(upars, dt_sim)
    # basic loglikelihood
    ll, Xt = rodeo.inference.basic(
        key=None,  # immaterial, since not used
        # ode specification
        ode_fun=fitz_fun,
        ode_weight=W,
        ode_init=X0,
        t_min=t_min,
        t_max=t_max,
        theta=theta,
        # solver parameters
        n_steps=n_steps,
        interrogate=rodeo.interrogate.interrogate_kramer,
        prior_weight=prior_Q,
        prior_var=prior_R,
        # observations
        obs_data=Y,
        obs_times=obs_times,
        obs_loglik=fitz_loglik,
        kalman_type="standard",
    )
    return -(ll + fitz_logprior(upars))


# keys
basic_key, fenrir_key, hmc_key, mcmc_key, magi_key = jax.random.split(key, num=5)

# starting values for optimization
upars_init = jnp.append(jnp.ones(len(theta)), X0[:, 0])
upars_init = jnp.append(upars_init, jnp.log(0.01 * jnp.ones(n_vars)))
# samples for posterior
n_samples = 100_000

# Laplace approximation via basic likelihood
Theta_basic = fitz_laplace(basic_key, fitz_neglogpost_basic, n_samples, upars_init)
\end{lstlisting}

\subsection[MCMC with BlackJAX]{MCMC with \bjax} \label{sec:mcmcbjax}

For the Basic, \fenrir, and \dalton methods, we can also use MCMC to sample from the closed-form posterior approximation $\hat{p}(\TTh \mid \YY_{0:M})$, when the Laplace approximation is deemed to be insufficiently accurate.
This can be easily and efficiently accomplished using the \bjax package\citep{blackjax}.  Like \rodeo, \bjax is written with \jax and offers many built-in MCMC algorithms to choose from.   Here we use the Hybrid Monte Carlo (HMC) algorithm~\citep{duane.etal87}, using \jax to conveniently compute the required gradient of $\log \hat p(\TTh \mid \YY_{0:M})$ via automatic differentiation.


Continuing with the example of the FN model introduced in Section~\ref{sec:laplaceapp}, the code below shows the basic usage of \bjax with HMC to sample from the parameter posterior $p(\tth \mid \YY_{0:M})$ using the Basic ODE likelihood approximation.  We use five HMC leapfrog steps and tune the remaining HMC parameters (step size and mass matrix) via the window adaptation algorithm described in~\cite{stan}. 
\begin{lstlisting}[language=iPython,firstline=225,lastline=286, caption={Parameter posterior for the FN model using \bjax HMC with the Basic ODE likelihood approximation.}]
# --- Import libraries and modules ---------------------------------------

import jax
import jax.numpy as jnp
import blackjax  # mcmc algorithms written in jax

# --- parameter inference: basic + hmc -----------------------------------


# blackjax uses the loglikelihood function instead of the negative
def fitz_logpost_basic(upars):
    return -fitz_neglogpost_basic(upars)


def fitz_hmc(key, logpost, n_samples, upars_init, num_steps):
    """
    Sample from the parameter posterior via the Hamiltonian Monte Carlo
    method for the FN model.

    Args:
        key : PRNG key.
        logpost : Function specifying the log-posterior distribution
                  n terms of the unconstrained parameter vector ``upars``.
        upars_init : Initial value to the optimization algorithm over ``logpost()``.
        n_samples : Number of posterior samples to draw.
        num_steps : Number of HMC leapfrog steps.

    Returns:
        JAX array of shape ``(n_samples, 5)`` of posterior samples from ``(theta, x0)``.
    """
    key, *subkeys = jax.random.split(key, num=3)
    # Stan window adaptation algorithm to start with reasonable parameters
    warmup = blackjax.window_adaptation(
        blackjax.hmc, logpost, num_integration_steps=num_steps
    )
    (initial_state, parameters), _ = warmup.run(subkeys[0], upars_init)
    kernel = blackjax.hmc(logpost, **parameters).step

    # standard in blackjax to write an inference loop for sampling
    def inference_loop(key, kernel, initial_state, n_samples):

        def one_step(state, rng_key):
            state, _ = kernel(rng_key, state)
            return state, state

        keys = jax.random.split(key, n_samples)
        # scan will automatically jit compile for you so no need to do so explicitly
        _, states = jax.lax.scan(one_step, initial_state, keys)
        return states

    uode_sample = inference_loop(subkeys[1], kernel, initial_state, n_samples).position
    if isinstance(uode_sample, dict):
        uode_sample = uode_sample["upars"]
    else:
        uode_sample = uode_sample[:, :5]
    # convert back to original scale
    ode_sample = uode_sample.at[:, :3].set(jnp.exp(uode_sample[:, :3]))
    return ode_sample


# HMC via basic likelihood
Theta_hmc = fitz_hmc(hmc_key, fitz_logpost_basic, n_samples, upars_init, num_steps=5)
\end{lstlisting}

\subsection{Efficient Blocking for Gaussian Likelihoods} \label{sec:effblock}

In the examples above, the Gaussian measurement model~\eqref{eq:fitznoise} was implemented by hand in the function \fct{fitz\_loglik}.  In contrast, the \fenrir and Gaussian \dalton algorithms are specifically designed for Gaussian noise models.  As discussed in Sections~\ref{sec:fenrir} and~\ref{sec:dalton}, these algorithms can benefit from blocking when the parameters $\DD_i^{(\pph)}$ and $\OOm_i^{(\pph)}$ of the noise model~\eqref{eq:measnorm} are block diagonal. For the FN example, we have $d=2$ ODE variables, $p=2$ derivatives in the IBM prior, and $s=1$ measurements per ODE variable, such that in \pkg{NumPy}/\jax array notation, the terms of~\eqref{eq:measnorm} corresponding to the measurement model~\eqref{eq:fitznoise} are given by
\begin{equation}\label{eq:fitzarray}
    \begin{aligned}
        \YY_i &= \begin{bmatrix}\begin{bmatrix}Y_i^{(V)}\end{bmatrix} &  \begin{bmatrix}Y_i^{(R)}\end{bmatrix}\end{bmatrix}, \\
        \DD_i^{(\pph)} &= \begin{bmatrix}\begin{bmatrix}\begin{bmatrix}1 & 0 & 0\end{bmatrix}\end{bmatrix},\begin{bmatrix}\begin{bmatrix}1 & 0 & 0\end{bmatrix}\end{bmatrix}\end{bmatrix}, \\
        \OOm_i^{(\pph)} &= \begin{bmatrix}\begin{bmatrix}\begin{bmatrix}\phi^2\end{bmatrix}\end{bmatrix},\begin{bmatrix}\begin{bmatrix}\phi^2\end{bmatrix}\end{bmatrix}\end{bmatrix}.
    \end{aligned}
\end{equation}
The code below shows how to construct the ODE likelihood approximation for the FN model using \fenrir as described in Section~\ref{sec:fenrir}, which could then be used represent the corresponding posterior $\hat{p}(\TTh \mid \YY_{0:M})$ via the Laplace approximation or via MCMC, as shown in the examples above.
\begin{lstlisting}[language=iPython,firstline=288,lastline=336, caption={Parameter posterior for the FN model using the Laplace approximation with \fenrir.}]
# --- Import libraries and modules -------------------------------------------

import jax
import jax.numpy as jnp
import jaxopt  # optimization library for finding mode in Laplace approximation
import rodeo.interrogate  # interrogation methods
import rodeo.inference  # parameter inference methods

# --- parameter inference: fenrir + laplace ----------------------------------

# gaussian measurement model specification in blocked form
n_meas = 1  # number of measurements per variable in obs_data_i
obs_data = jnp.expand_dims(Y, axis=-1)
obs_weight = jnp.zeros((len(obs_data), n_vars, n_meas, n_deriv))
obs_weight = obs_weight.at[:].set(jnp.array([[[1.0, 0.0, 0.0]], [[1.0, 0.0, 0.0]]]))
obs_var = jnp.zeros((len(obs_data), n_vars, n_meas, n_meas))
obs_var = obs_var.at[:].set(noise_sd**2 * jnp.array([[[1.0]], [[1.0]]]))


def fitz_neglogpost_fenrir(upars):
    "Negative logposterior for basic approximation."
    theta, X0, prior_Q, prior_R = fitz_constrain_pars(upars, dt_sim)
    # fenrir loglikelihood
    ll = rodeo.inference.fenrir(
        key=None,  # immaterial, since not used
        # ode specification
        ode_fun=fitz_fun,
        ode_weight=W,
        ode_init=X0,
        t_min=t_min,
        t_max=t_max,
        theta=theta,
        # solver
        n_steps=n_steps,
        interrogate=rodeo.interrogate.interrogate_kramer,
        prior_weight=prior_Q,
        prior_var=prior_R,
        # gaussian measurement model
        obs_data=obs_data,
        obs_times=obs_times,
        obs_weight=obs_weight,
        obs_var=obs_var,
        kalman_type="standard",
    )
    return -(ll + fitz_logprior(upars))


# Laplace approximation via fenrir likelihood
Theta_fenrir = fitz_laplace(fenrir_key, fitz_neglogpost_fenrir, n_samples, upars_init)
\end{lstlisting}

\subsection{Marginal MCMC}

The marginal MCMC method shares the same API as the \bjax MCMC described in Section~\ref{sec:mcmcbjax}. Referring back to Algorithm~\ref{alg:Chkrebtii}, the first step is to choose a proposal distribution on Line~\ref{ln:rwkernel}. For this, we use a random walk (RW) kernel:
\begin{equation}\label{eq:rwkernel}
    \TTh^\prop \mid \TTh^\curr \sim \N(\TTh^\curr, \diag(\ssi_{rw}^2)),
\end{equation}
where $\diag(\ssi_{rw}^2)$ is a tuning parameter for the MCMC algorithm. While \bjax provides an RW-MCMC sampler, it does not support a Metropolis–Hastings acceptance ratio that depends on the auxiliary random variable $\XX_{1:N}$, as required for pseudo-marginal MCMC. Therefore, we use the \bjax API to define a pseudo-marginal MCMC sampler with an RW kernel, which we provide in the \code{rodeo.random_walk_aux} module.

There are three main steps to using the marginal MCMC method. First, a likelihood function must be defined, where the ODE solver uses the interrogation method of~\citep{chkrebtii.etal16} to sample from the solution posterior $\hat{p}_\L(\XX_{1:N}, \ipar_{1:N} \mid \ZZ_{1:N} = \bz, \tth, \eet)$. Second, a kernel must be defined to execute Algorithm~\ref{alg:Chkrebtii}, for which we use the RW kernel in~\eqref{eq:rwkernel}. Finally, an inference loop, similar to the one in Section~\ref{sec:mcmcbjax}, is used to draw MCMC samples from the parameter posterior.

The code below shows how to use the marginal MCMC method to sample from the parameter posterior $p(\tth \mid \YY_{0:M})$ for the FN model. We tune the hyperparameter in~\eqref{eq:rwkernel} to achieve an acceptance rate between 20\% and 30\%.
\begin{lstlisting}[language=iPython,firstline=338, lastline=425,caption={Parameter posterior for the FN model using the \chkrebtii marginal MCMC method.}]
# --- Import libraries and modules ---------------------------------------

import jax
import jax.numpy as jnp
import functools  # partial to fix kalman_type
import rodeo  # ODE solvers
import rodeo.interrogate  # interrogation methods
import rodeo.inference  # parameter inference methods
from rodeo.inference import random_walk_aux  # random walk kernel for marginal mcmc

# --- parameter inference: chkrebtii -------------------------------------

# fitx the value for kalman_type
interrogate_chkbretii_partial = functools.partial(
    rodeo.interrogate.interrogate_chkrebtii, kalman_type="standard"
)


def fitz_logpost_mcmc(key, upars):
    """
    Computes marginal MCMC posterior.

    Also returns solution path Xt that was generated.
    """
    theta, X0, prior_Q, prior_R = fitz_constrain_pars(upars, dt_sim)
    Xt = rodeo.solve_sim(
        key=key,
        # define ode
        ode_fun=fitz_fun,
        ode_weight=W,
        ode_init=X0,
        t_min=t_min,
        t_max=t_max,
        theta=theta,
        # solver parameters
        n_steps=n_steps,
        interrogate=interrogate_chkbretii_partial,
        prior_weight=prior_Q,
        prior_var=prior_R,
        kalman_type="standard",
    )
    ode_data = Xt[obs_ind]
    lp = fitz_logprior(upars) + fitz_loglik(Y, ode_data)
    return lp, Xt


def fitz_mcmc(key, n_samples, upars):
    r"""
    Marginal MCMC using blackjax.
    """
    key, subkey = jax.random.split(key)
    # initialize mcmc
    state = random_walk_aux.init(
        position=upars, logdensity_fn=lambda upars: fitz_logpost_mcmc(subkey, upars)
    )
    keys = jax.random.split(key, num=n_samples)
    # blackjax
    # RW with auxiliary outputs
    rwa_kernel = random_walk_aux.build_additive_step()
    # standard deviation of the random walk
    scale = jnp.array([0.01, 0.1, 0.01, 0.01, 0.01, 0.01, 0.01])

    # inference loop
    def marginal_rw_step(key, state, sigma):
        "One step of Marginal Random Walk."
        keys = jax.random.split(key, num=2)
        return rwa_kernel(
            rng_key=keys[0],  # for RW proposal
            state=state,
            # logpost for each step has its own key for generating Xt
            logdensity_fn=lambda position: fitz_logpost_mcmc(keys[1], position),
            random_step=blackjax.mcmc.random_walk.normal(sigma=sigma),
        )

    def step(state, rng_key):
        state, info = marginal_rw_step(key=rng_key, state=state, sigma=scale)
        return state, (state, info)

    # scan will automatically jit compile for you so no need to do so explicitly
    _, out = jax.lax.scan(f=step, init=state, xs=keys)
    uode_sample = out[0].position[:, :5]
    # convert back to original scale
    ode_sample = uode_sample.at[:, :3].set(jnp.exp(uode_sample[:, :3]))
    return ode_sample


# Chkrebtii marginal MCMC
Theta_mcmc = fitz_mcmc(mcmc_key, n_samples, upars_init)
\end{lstlisting}

\subsection{MAGI} \label{sec:magiex}

The \magi method requires a few different inputs compared to previous inference methods. Since the $\beta$-modulated \magi posterior in~\eqref{eq:magimodbeta} is $p_\beta(\TTh, \tilde \UU_{1:N} \mid \YY_{0:M}, \ZZ_{1:N} = \bz)$, the log density function must include both the parameter $\TTh$ and the latent process $\tilde \UU_{1:N}$. Additionally, a function is needed to apply the ODE function $\ff_\tth(\tilde \UU_n, t_n)$ at every time point. Finally, \magi requires the specification of the number of active derivatives to accurately identify the subset described in Section~\ref{sec:magi}, where we assume that higher-order derivatives do not directly affect the ODE and are included solely for a smoother solution.

For inference, we follow the same steps as the original \magi method in~\citep{yang.etal21}. The hyperparameter $\eet$ for the Markov prior is assumed to remain fixed during MCMC. Our experiments indicate an insensitivity to the choice of $\eet$ over orders of magnitude.  Thus, for simplicity we set $\eet = 0.1$. The code below demonstrates the basic usage of HMC from \bjax to sample from the joint posterior $p_\beta(\tth, \tilde \UU_{1:N} \mid \YY_{0:M}, \ZZ_{1:N} = \bz)$ using the $\beta$-modulated \magi in~\eqref{eq:magimodbeta}. We choose the prior temperature $\beta = \eet^{-2} \dt^{2-2q} \dt'$ where $\dt$ is the discretization step size and $\dt'$ is the interobservation step size. We choose $\beta$ in this manner so that it is proportional to $R_{11}^{-1}$ from the IBM prior in Section~\ref{sec:prior}, scaled by the factor $\dt'/\dt$ as suggested by~\cite{yang.etal21}. We use 200 leapfrog steps and tune the remaining HMC parameters via the window adaptation algorithm~\citep{stan}. For numerical stability, we use the square-root Kalman filter.
\begin{lstlisting}[language=iPython,firstline=427,caption={Parameter posterior for the FN model using \bjax HMC with the \magi method.}]
# --- Import libraries and modules ---------------------------------------

import jax
import jax.numpy as jnp
import blackjax  # mcmc algorithms written in jax
import rodeo.prior  # IBM prior
import rodeo.inference  # parameter inference methods

# --- parameter inference: magi ------------------------------------------


def fitz_expand(X, theta):
    """
    Expand FN data.
    """
    return jax.vmap(
        fun=lambda x, t: fitz_init_pad(x, t, theta=theta),
    )(X, sim_times)


def fitz_magi_logpost_fn(upars, ode_data_subset):
    """
    Joint posterior function for the MAGI approximation.
    """
    theta = jnp.exp(upars[:3])
    x0 = upars[3:5]
    ode_data_subset = ode_data_subset.at[0].set(x0)
    beta = (
        dt_obs * dt_sim ** (2 - 2 * n_deriv) * sigma[0] ** (-2)
    )  # magi prior temperature
    lp = (
        rodeo.inference.magi_logdens(
            ode_data_subset=ode_data_subset,
            ode_expand=fitz_expand,
            n_active=2,  # 0th and 1st order derivatives
            prior_weight=prior_Q,
            prior_var=prior_Rhalf,
            theta=theta,
            # use square-root version for numerical stability
            kalman_type="square-root",
        )
        / beta
    )
    sim_times = jnp.linspace(t_min, t_max, n_steps + 1)
    ode_data = ode_data_subset[jnp.searchsorted(sim_times, obs_times)][:, :, None]
    lp += fitz_loglik(Y, ode_data) + fitz_logprior(upars)
    return lp


fitz_magi_logpost = lambda x: fitz_magi_logpost_fn(**x)
sigma = jnp.array([0.1] * n_vars)  # fixed at arbitrary value
prior_Q, prior_R = rodeo.prior.ibm_init(dt=dt_sim, n_deriv=n_deriv, sigma=sigma)
prior_Rhalf = jax.vmap(jnp.linalg.cholesky)(prior_R)
upars_init = jnp.append(jnp.ones(len(theta)), X0[:, 0])
# linear interpolation as an initial guess for MCMC
ode_subset_init = jax.vmap(jnp.interp, in_axes=[None, None, 1])(
    sim_times, obs_times, Y
).T
magi_upars_init = {"upars": upars_init, "ode_data_subset": ode_subset_init}
Theta_magi = fitz_hmc(
    magi_key, fitz_magi_logpost, n_samples, magi_upars_init, num_steps=200
)
\end{lstlisting}

\newpage

\subsection{Comparison of Posterior Approximations} \label{sec:fncomp}

\begin{figure}[!htb]    \includegraphics[width=\linewidth]{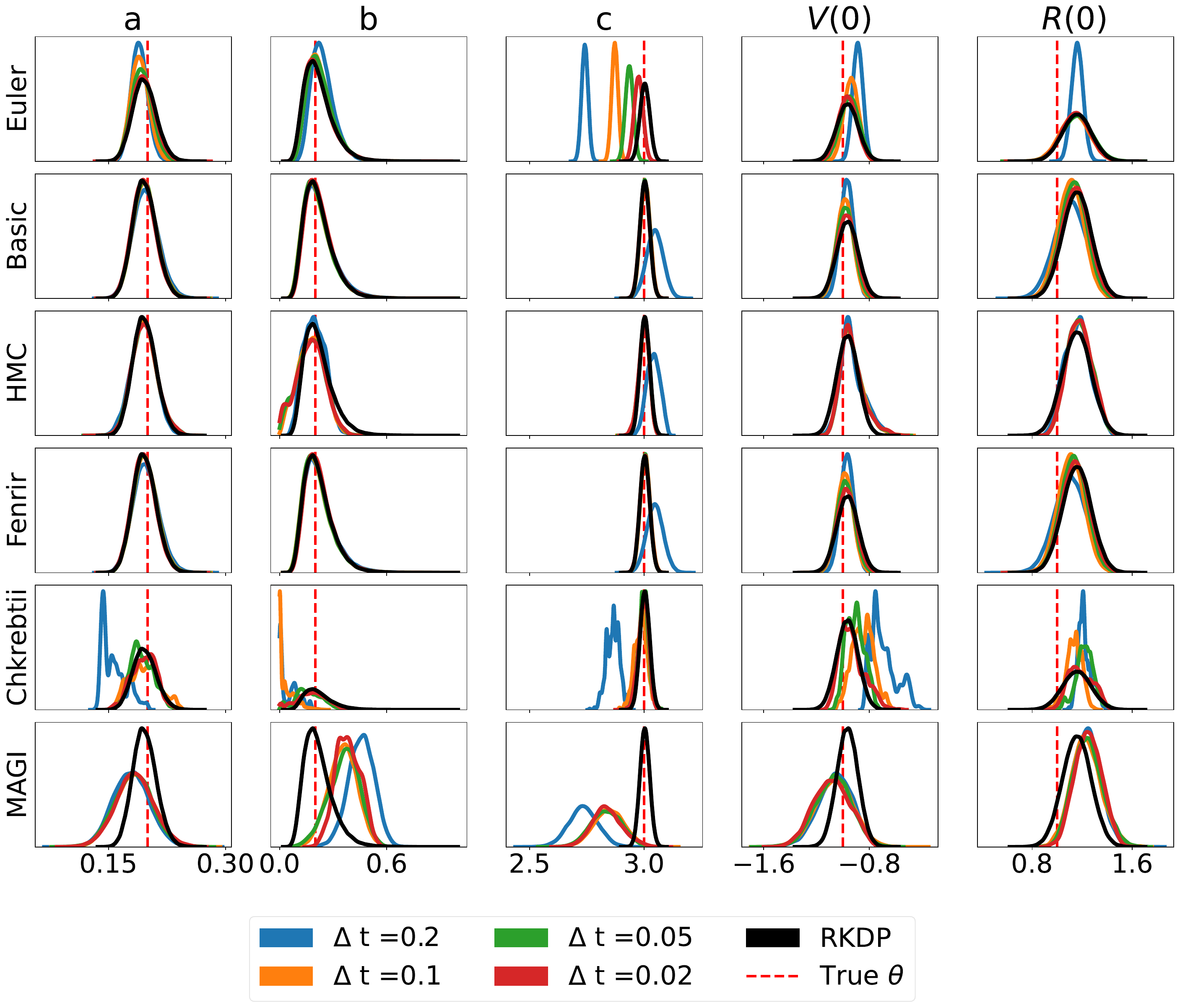}
    \caption{
      Parameter posteriors for the FN model at different step sizes $\dt$.  The Laplace approximation is used for the Euler ODE solver, the RKDP solver, the Basic ODE likelihood approximation, and \fenrir. Full posteriors are shown for the \chkrebtii approximation (marginal MCMC), the Basic likelihood approximation (HMC), and \magi (HMC).
    }
    \label{fig:fitzpost}
\end{figure}

Figure~\ref{fig:fitzpost} displays the parameter posteriors for various approximations discussed above at different solver step sizes $\dt = T/N$.  In addition to the five posterior approximations presented in Sections~\ref{sec:laplaceapp}-\ref{sec:magiex},
also included are Laplace posteriors for the basic approximation method using (i) an Euler solver and (ii) a highly accurate deterministic solver,
namely, a Runge-Kutta 5(4) solver with Dormand-Prince step size adaptation~\citep{dormand.prince80} -- RKDP -- of which the implementation is provided by the \diffrax library~\citep{kidger21}. We shall assume that the output of the RKDP solver differs negligibly from the true ODE solution.

For parameter $c$, the Euler posteriors do not get close to those of RKDP before $\dt = 0.02$.  In contrast, the Basic approximation and \fenrir posteriors are very close to the RKDP posterior as of $\dt = 0.1$. The \chkrebtii marginal MCMC method covers the true parameter values at $\dt \leq 0.05$ whereas the \bjax MCMC with the Basic algorithm does so at all step sizes. The \magi MCMC parameter posteriors cover the true parameter values at $\dt \leq 0.1$.


\subsection{The Hes1 ODE Model} \label{sec:hes1}

The Hes1 model~\citep{yang.etal21} is a three-variable dynamic system $\xx(t) = (P(t), M(t), H(t))$ that describes the oscillation of the Hes1-protein (P) and Hes1-mRNA (M) with an Hes1-interacting (H) factor acting as a stabilizer. The log-scale Hes1 model is
\begin{equation} \label{eq:loghes1}
    \begin{aligned}
        \dv{\log P(t)}{t} &= -aH(t) + bM(t)/P(t) - c, \\
        \dv{\log M(t)}{t} &= -d + \frac{e}{1+P(t)^2M(t)}, \\
        \dv{\log H(t)}{t} &= -aP(t) + \frac{f}{1+P(t)^2H(t)} - g.
    \end{aligned}
\end{equation}
It contains 10 parameters $\tth = (a, b, c, d, e, f, g, P(0), M(0), H(0))$ with $\tth > 
\bz$.  In this case, the measurements consist of noisy observations of only $P(t)$ and $M(t)$, with $H(t)$ completely unobserved.  Moreover, $P(t)$ and $M(t)$ measured at different times; each is measured at 15-minute intervals for a four-hour period, but $P(t)$ measurements start at time $t=0$ whereas $M(t)$ measurements start at time $t=7.5$ minutes. Following~\cite{yang.etal21}, the measurement error model is
\begin{equation}\label{eq:hes1noise}
    Y^{(K)}_i \ind \N(\log K(t^{(K)}_i), \phi^2), \qquad K \in \{P, M\},
\end{equation}
where $\phi = 0.15$.
Data is simulated in the same way as Section~\ref{sec:laplaceapp} for $Y_i$ with true parameter values $\tth = (0.022, 0.3, 0.031, 0.028, 0.5, 20, 0.3, 1.439, 2.037, 17.904)$.

\subsection{Efficient Blocking with Unobserved Components} \label{sec:effblockuc}
We define $ \YY_i, \DD^{(\pph)}_i$ and $\OOm^{(\pph)}_i$ in a similar manner to Section~\ref{sec:effblock}. The caveat here is that at each observation time point $t'_i$ we have $H(t'_i)$ unobserved and either $P(t'_i)$ or $M(t'_i)$ unobserved. For each unobserved component $K$, the observation $Y_i^{(K)}$ will be zero and the matrices $\DD^{(\pph)}_i$ and $\OOm^{(\pph)}_i$ will have zeros in the row corresponding to the unobserved component. This ensures Algorithm~\ref{alg:Fenrir} and~\ref{alg:dalton} add zero whenever there is an unobserved component (i.e., the log density of $\N(0;0,0)$). The code below shows how to construct the ODE likelihood approximation for the Hes1 model using \dalton as described in Section~\ref{sec:dalton}. 
\begin{lstlisting}[language=iPython, caption={Parameter posterior for the Hes1 model using the Laplace approximation with \dalton where components are partially observed.}]
# --- 0. Import libraries and modules ------------------------------------

import jax
import jax.numpy as jnp
import jax.scipy as jsp
import jaxopt
import rodeo
import rodeo.prior
import rodeo.interrogate
import rodeo.inference
from rodeo.utils import first_order_pad

# --- 1. Define the ODE-IVP ----------------------------------------------


def hes1_fun(X, t, theta):
    "Hes1 on the log-scale in rodeo format."
    P, M, H = jnp.exp(X[:, 0])
    a, b, c, d, e, f, g = theta
    logP = -a * H + b * M / P - c
    logM = -d + e / (1 + P * P) / M
    logH = -a * P + f / (1 + P * P) / H - g
    return jnp.array([[logP], [logM], [logH]])


n_vars = 3
n_deriv = 3

# initial value for the ODE-IVP on log-scale
x0 = jnp.log(jnp.array([1.439, 2.037, 17.904]))
theta = jnp.array([0.022, 0.3, 0.031, 0.028, 0.5, 20, 0.3])  # ODE parameters
W, hes1_init_pad = first_order_pad(hes1_fun, n_vars, n_deriv)

# Time interval on which a solution is sought.
t_min = 0.0
t_max = 240.0

# --- Define the prior process -------------------------------------------

# IBM process scale factor
sigma = jnp.array([0.1] * n_vars)

# simulation times
dt_sim = 0.375
n_steps = int((t_max - t_min) / dt_sim)

# --- parameter inference: dalton ----------------------------------


def hes1_logprior(upars):
    "Logprior on unconstrained model parameters."
    n_theta = 10  # number of ODE + IV parameters
    lpi = jax.scipy.stats.norm.logpdf(x=upars[:n_theta], loc=0.0, scale=10.0)
    return jnp.sum(lpi)


def hes1_constrain_pars(upars, dt):
    """
    Convert unconstrained optimization parameters into rodeo inputs.

    Args:
        upars: Parameters vector on unconstrainted scale.
        dt: Discretization grid size.

    Returns:
        tuple with elements:
        - theta : ODE parameters.
        - X0 : Initial values in rodeo format.
        - Q, R : Prior matrices.
    """
    theta = jnp.exp(upars[:7])
    x0 = upars[7:10]
    X0 = hes1_init_pad(x0, 0, theta=theta)
    sigma = jnp.exp(upars[10:])
    Q, R = rodeo.prior.ibm_init(dt=dt, n_deriv=n_deriv, sigma=sigma)
    return theta, X0, Q, R


# Produce a Pseudo-RNG key
key = jax.random.PRNGKey(0)

# gaussian measurement model specification in blocked form
obs_data = jnp.load("hes1data.npy")  # load the data
noise_sd = 0.15  # Standard deviation in noise model
n_steps_obs = obs_data.shape[0]
# observation times
obs_times = jnp.linspace(t_min, t_max, num=n_steps_obs)
n_meas = 1  # number of measurements per variable in Y_i
obs_weight = jnp.zeros((n_steps_obs, n_vars, n_meas, n_deriv))
obs_weight = obs_weight.at[::2, 0].set(jnp.array([[[1.0, 0.0, 0.0]]]))
obs_weight = obs_weight.at[1::2, 1].set(jnp.array([[[1.0, 0.0, 0.0]]]))
obs_var = jnp.zeros((n_steps_obs, n_vars, n_meas, n_meas))
obs_var = obs_var.at[::2, 0].set(noise_sd**2 * jnp.array([[[1.0]]]))
obs_var = obs_var.at[1::2, 1].set(noise_sd**2 * jnp.array([[[1.0]]]))


def hes1_neglogpost_dalton(upars):
    "Negative logposterior for dalton approximation."
    theta, X0, prior_Q, prior_R = hes1_constrain_pars(upars, dt_sim)
    # Gaussian DALTON loglikelihood
    ll = rodeo.inference.dalton(
        key=None,  # immaterial, since not used
        # ode specification
        ode_fun=hes1_fun,
        ode_weight=W,
        ode_init=X0,
        t_min=t_min,
        t_max=t_max,
        theta=theta,
        # solver
        n_steps=n_steps,
        interrogate=rodeo.interrogate.interrogate_kramer,
        prior_weight=prior_Q,
        prior_var=prior_R,
        kalman_type="standard",
        # gaussian measurement model
        obs_data=obs_data,
        obs_times=obs_times,
        obs_weight=obs_weight,
        obs_var=obs_var,
    )
    return -(ll + hes1_logprior(upars))


def hes1_laplace(key, neglogpost, n_samples, upars_init):
    """
    Sample from the Laplace approximation to the parameter posterior for the hes1 model.

    Args:
        key : PRNG key.
        neglogpost: Function specifying the negative log-posterior distribution
                    in terms of the unconstrained parameter vector ``upars``.
        upars_init: Initial value to the optimization algorithm over ``neglogpost()``.
        n_samples : Number of posterior samples to draw.

    Returns:
        JAX array of shape ``(n_samples, )`` of posterior samples from ``(theta, x0)``.
    """
    n_theta = 10  # number of ODE + IV parameters
    # find mode of neglogpost()
    solver = jaxopt.ScipyMinimize(fun=neglogpost, method="Newton-CG", jit=True)
    opt_res = solver.run(upars_init)
    upars_mean = opt_res.params
    # variance estimate
    upars_fisher = jax.jacfwd(jax.jacrev(neglogpost))(upars_mean)
    # unconstrained ode+iv parameter variance estimate
    uode_var = jsp.linalg.inv(upars_fisher[:n_theta, :n_theta])
    uode_mean = upars_mean[:n_theta]
    # sample from Laplace approximation
    uode_sample = jax.random.multivariate_normal(
        key=key, mean=uode_mean, cov=uode_var, shape=(n_samples,)
    )
    # convert back to original scale
    ode_sample = jnp.exp(uode_sample)
    return ode_sample


# Laplace approximation with Dalton
# starting values for optimization
upars_init = jnp.append(-2 * jnp.ones(len(theta)), x0)
upars_init = jnp.append(upars_init, jnp.log(1e-1 * jnp.ones(n_vars)))
# samples for posterior
n_samples = 100_000
Theta_dalton = hes1_laplace(key, hes1_neglogpost_dalton, n_samples, upars_init)
\end{lstlisting}

\begin{figure}[!htb]
    \centering
    \includegraphics[width=.95\linewidth]{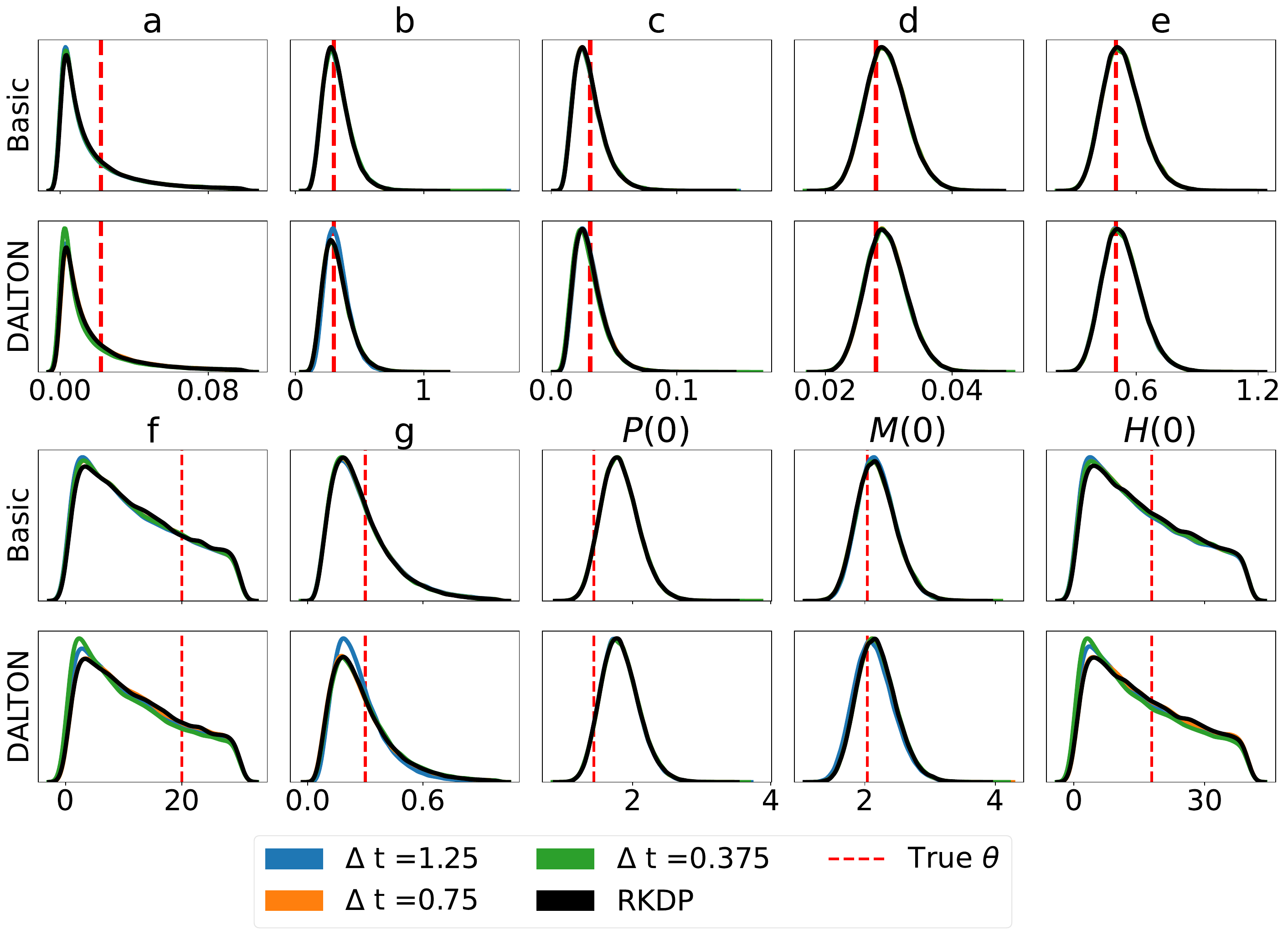}
    \caption{
        The parameter posteriors of the Basic and \dalton approximations for the Hes1 model.
    }
    \label{fig:hes1post}
\end{figure}

Figure~\ref{fig:hes1post} displays the parameter posteriors of the Laplace approximation using the Basic and Gaussian \dalton algorithms. We were unable to obtain parameter estimates for solver step size $\dt > 2.0$ minutes due to numerical instability. However, the Basic and \dalton posteriors for $\dt \le 0.75$ minutes are almost indistinguishable from the true posterior.

\subsection{The SEIRAH Model} \label{sec:seirah}

The SEIRAH model is a six-compartment epidemiological model used to describe Covid-19 dynamics by~\cite{prague.etal20}. 
The six compartments of a given population are: $S$ susceptible, $E$ latent, $I$ ascertained infectious, $R$ removed (both recovered and deceased), $A$ unascertained infectious, and $H$ hospitalized.  The evolution of these quantities according to the SEIRAH model is given by
\begin{equation} \label{eq:seirah}
    \begin{aligned}
        \dv{S(t)}{t} &= -\frac{bS(t)(I(t) + \alpha A(t))}{N}, &
        \dv{E(t)}{t} &= \frac{bS(t)(I(t) + \alpha A(t))}{N} - \frac{E(t)}{D_e}, \\
        \dv{I(t)}{t} &= \frac{rE(t)}{D_e} - \frac{I(t)}{D_q} - \frac{I(t)}{D_I}, &
        \dv{R(t)}{t} &= \frac{I(t) + A(t)}{D_I} + \frac{H_t}{D_h}, \\
        \dv{A(t)}{t} &= \frac{(1-r)E(t)}{D_e} - \frac{A(t)}{D_I}, &
        \dv{H(t)}{t} &= \frac{I(t)}{D_q} - \frac{H(t)}{D_h},
    \end{aligned}
\end{equation}
where $N = S(t) + E(t) + I(t) + R(t) + A(t) + H(t)$ is fixed and so is $D_h=30$~\citep{prague.etal20}.  Again, we only have (indirect) observations of two of the six compartments.  That is, let
\begin{equation}
    \begin{aligned}\label{eq:seirahobs}
        \tilde I(t) &= \frac{rE(t)}{D_e}, &
        \tilde H(t) &= \frac{I(t)}{D_q}
    \end{aligned}
\end{equation}
denote the new cases entering their respective compartments.  The measurement model of~\cite{prague.etal20} is
\begin{equation}\label{eq:seirahnoise}
    \begin{aligned}
        Y_i^{(\tilde K)} \ind \operatorname{Poisson}(\tilde K(t_i)), \qquad K \in \{I, H\},
    \end{aligned}
\end{equation}
where $t_i = i$ days with $i = 0, \ldots, 60$.  The unknown parameters are $\tth = (b, r, \alpha, D_e, D_I, D_q, E(0), I(0))$ with $\tth > 0$.  Data was simulated with true parameter values $\tth = (2.23, 0.034, 0.55, 5.1, 2.3, 1.13, 15492, 21752)$ and remaining initial values $S^{(0)}_0 = 63884630$, $R^{(0)}_0 = 0$, $A^{(0)}_0 = 618013$, and $H^{(0)}_0 = 13388$ (as reported by~\cite{prague.etal20}).

\subsection{Non-Gaussian DALTON} \label{sec:ngdalton}

The code below shows how to construct the ODE likelihood approximation for the SEIRAH model using non-Gaussian \dalton in Algorithm~\ref{alg:daltonng} as described in Appendix~\ref{sec:daltonng}. The unobserved components in~\eqref{eq:seirahnoise} are directly encoded in \code{obs_loglik_i}.

\begin{figure} [!htb]    \includegraphics[width=\linewidth]{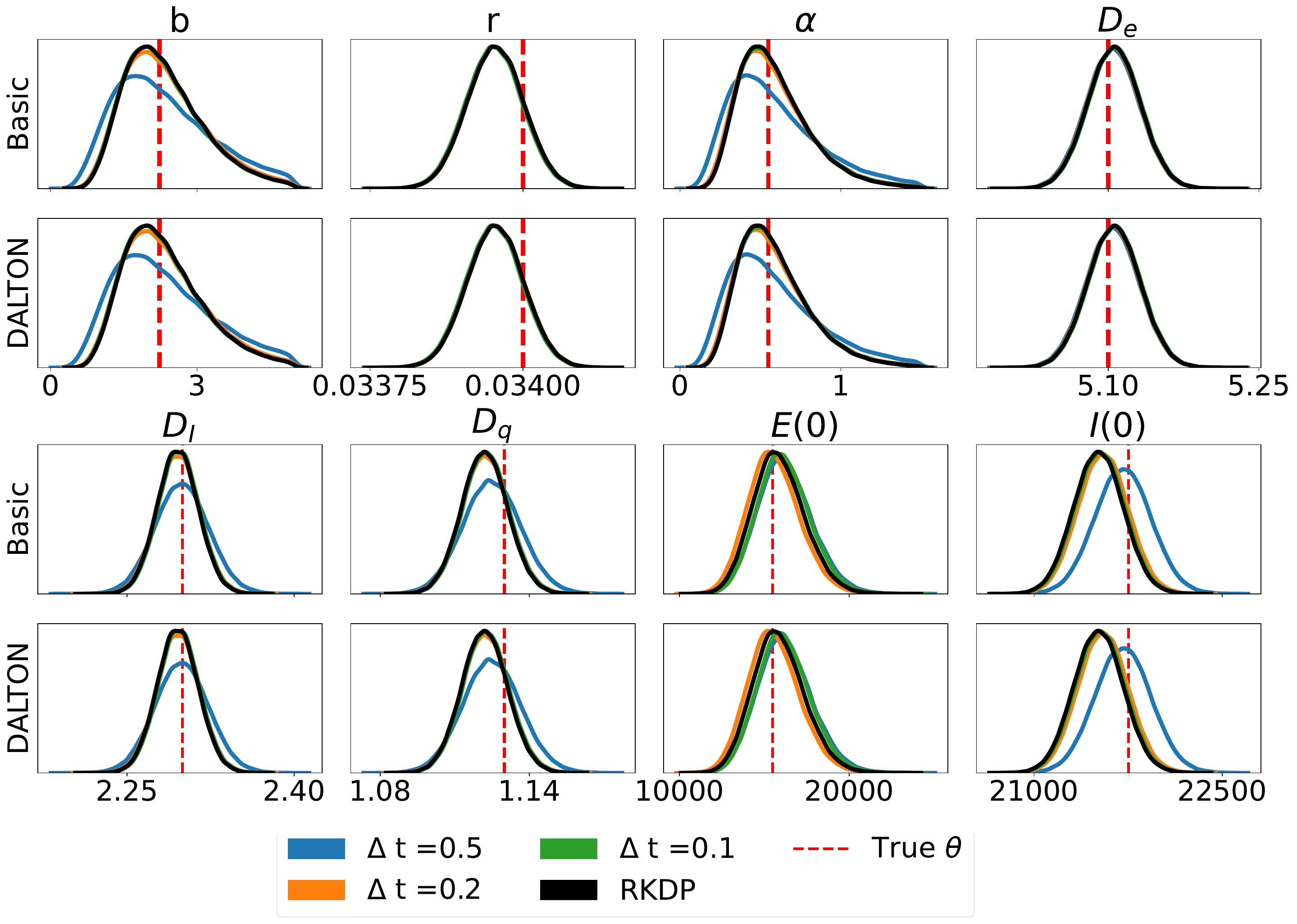}
    \caption{
        The parameter posteriors of the Basic and non-Gaussian \dalton methods for the SEIRAH model.
    }
    \label{fig:seirahpost}
\end{figure}

Figure~\ref{fig:seirahpost} displays the Basic and non-Gaussian \dalton posteriors for various step sizes $\dt$.  Inference for the Basic approximation was carried out largely the same way as described in Section~\ref{sec:laplaceapp}.  All posteriors cover the true value of $\tth$ and they are indistinguishable from the RKDP posterior at $\dt = 0.1$.

\begin{lstlisting}[language=iPython, caption={Parameter posterior for the SEIRAH model using the Laplace approximation with non-Gaussian \dalton.}]
# --- 0. Import libraries and modules ------------------------------------


import jax
import jax.numpy as jnp
import jax.scipy as jsp
import jaxopt
import rodeo
import rodeo.prior
import rodeo.interrogate
import rodeo.inference

# helper function to help with the padding in rodeo format
from rodeo.utils import first_order_pad

# --- 1. Define the ODE-IVP ----------------------------------------------


def seirah_fun(X, t, theta):
    "SEIRAH ODE in rodeo format."
    S, E, I, R, A, H = X[:, 0]
    N = S + E + I + R + A + H
    b, r, alpha, D_e, D_I, D_q = theta
    D_h = 30
    dS = -b * S * (I + alpha * A) / N
    dE = b * S * (I + alpha * A) / N - E / D_e
    dI = r * E / D_e - I / D_q - I / D_I
    dR = (I + A) / D_I + H / D_h
    dA = (1 - r) * E / D_e - A / D_I
    dH = I / D_q - H / D_h
    return jnp.array([[dS], [dE], [dI], [dR], [dA], [dH]])


def seirah_init(pars):
    "SEIRAH initial values in rodeo format."
    x0 = jnp.array([63884630.0, jnp.nan, jnp.nan, 0.0, 618013.0, 13388.0])
    # only two of the initial values are part of the inference
    x0 = x0.at[1:3].set(pars[6:8])
    return seirah_init_pad(x0, 0, theta=pars[:6])

n_deriv = 3
n_vars = 6
W, seirah_init_pad = first_order_pad(seirah_fun, n_vars, n_deriv)
theta = jnp.array([2.23, 0.034, 0.55, 5.1, 2.3, 1.13])  # ODE parameters

# Time interval on which a solution is sought.
t_min = 0.0
t_max = 60.0

# --- Define the prior process -------------------------------------------

# IBM process scale factor
sigma = jnp.array([0.1] * n_vars)

# simulation times
dt_sim = 0.02
n_steps = int((t_max - t_min) / dt_sim)

# --- parameter inference: daltonng ----------------------------


def seirah_logprior(upars):
    "Logprior on unconstrained model parameters."
    n_theta = 8  # number of ODE + IV parameters
    lpi = jax.scipy.stats.norm.logpdf(x=upars[:n_theta], loc=0.0, scale=10.0)
    return jnp.sum(lpi)


def seirah_constrain_pars(upars, dt):
    """
    Convert unconstrained optimization parameters into rodeo inputs.

    Args:
        upars : Parameters vector on unconstrainted scale.
        dt : Discretization grid size.

    Returns:
        tuple with elements:
        - theta : ODE parameters.
        - X0 : Initial values in rodeo format.
        - Q, R : Prior matrices.
    """
    pars = jnp.exp(upars)
    theta = pars[:6]
    X0 = seirah_init(pars)
    Q, R = rodeo.prior.ibm_init(dt=dt, n_deriv=n_deriv, sigma=pars[8:])
    return theta, X0, Q, R


# Produce a Pseudo-RNG key
key = jax.random.PRNGKey(0)

# non-gaussian measurement model
obs_data = jnp.load("seirahdata.npy")  # load the data
n_steps_obs = obs_data.shape[0]
# observation times
obs_times = jnp.linspace(t_min, t_max, num=n_steps_obs)


def seirah_obs_loglik_i(obs_data_i, ode_data_i, ind, theta):
    """
    Likelihood function for the SEIRAH model in non-Gaussian DALTON format.

    Args:
        obs_data_i : Observations at index ``ind``.
        ode_data_i :  ODE solution at index ``ind``.
        ind : Index to determine the observation loglikelihood function.

    Returns:
        Loglikelihood of ``obs_data_i`` at index ``ind``.
    """
    b, r, alpha, D_e, D_I, D_q = theta
    I_in = r * ode_data_i[1, 0] / D_e
    H_in = ode_data_i[2, 0] / D_q
    Xin = jnp.array([I_in, H_in])
    return jnp.sum(jax.scipy.stats.poisson.logpmf(obs_data_i.flatten(), Xin))


def seirah_neglogpost_daltonng(upars):
    "Negative logposterior for non-Gaussian DALTON approximation."
    theta, X0, prior_Q, prior_R = seirah_constrain_pars(upars, dt_sim)
    # non-Gaussian DALTON loglikelihood
    ll = rodeo.inference.daltonng(
        key=None,  # immaterial, since not used
        # ode specification
        ode_fun=seirah_fun,
        ode_weight=W,
        ode_init=X0,
        t_min=t_min,
        t_max=t_max,
        theta=theta,
        # solver
        n_steps=n_steps,
        interrogate=rodeo.interrogate.interrogate_kramer,
        prior_weight=prior_Q,
        prior_var=prior_R,
        # non-gaussian measurement model
        obs_data=obs_data,
        obs_times=obs_times,
        obs_loglik_i=seirah_obs_loglik_i,
        kalman_type="standard",
    )
    return -(ll + seirah_logprior(upars))


def seirah_laplace(key, neglogpost, n_samples, upars_init):
    """
    Sample from the Laplace approximation to the parameter posterior for the
    SEIRAH model.

    Args:
        key : PRNG key.
        neglogpost: Function specifying the negative log-posterior distribution
                    in terms of the unconstrained parameter vector ``upars``.
        upars_init: Initial value to the optimization algorithm over ``neglogpost()``.
        n_samples : Number of posterior samples to draw.

    Returns:
        JAX array of shape ``(n_samples, )`` of posterior samples from ``(theta, x0)``.
    """
    n_theta = 8  # number of ODE + IV parameters
    # find mode of neglogpost()
    solver = jaxopt.ScipyMinimize(fun=neglogpost, method="Newton-CG", jit=True)
    opt_res = solver.run(upars_init)
    upars_mean = opt_res.params
    # variance estimate
    upars_fisher = jax.jacfwd(jax.jacrev(neglogpost))(upars_mean)
    # unconstrained ode+iv parameter variance estimate
    uode_var = jsp.linalg.inv(upars_fisher[:n_theta, :n_theta])
    uode_mean = upars_mean[:n_theta]
    # sample from Laplace approximation
    uode_sample = jax.random.multivariate_normal(
        key=key, mean=uode_mean, cov=uode_var, shape=(n_samples,)
    )
    # convert back to original scale
    ode_sample = jnp.exp(uode_sample)
    return ode_sample


# Laplace approximation with non-Gaussian Dalton
# starting values for optimization
upars_init = jnp.append(jnp.log(theta), jnp.log(jnp.array([15492, 21752])))
upars_init = jnp.append(upars_init, jnp.log(0.01 * jnp.ones(n_vars)))
# samples for posterior
n_samples = 100_000
Theta_dalton = seirah_laplace(key, seirah_neglogpost_daltonng, n_samples, upars_init)
\end{lstlisting}

\subsection[Speed Comparisons for rodeo Solver]{Speed Comparisons for \rodeo Solver}

Table~\ref{tab:timings} displays the speed of \rodeo relative to different ODE solvers for the four examples above.
\begin{table}[!htb]
    \caption{Speed of \rodeo relative to other ODE solvers.} \label{tab:timings}
    \centering
    \begin{tabular}{llllllp{4.2em}}
        \toprule
      Model & $d$ & $d \cdot p$ & $N$ & LSODA & RKDP & \rodeo (no blocking) \\ 
        \midrule
        Chkrebtii & 1 & 4 & 30 & 4.8 & 12.64 & 0.97 \\ 
        FN & 2 & 6 & 250 & 12.93 & 3.94 & 3.49 \\ 
        Hes1 & 3 & 9 & 120 & 3.43 & 3.53 & 2.77 \\ 
        SEIRAH & 6 & 18 & 80 & 2.28 & 2.86 & 4.42 \\ 
        \bottomrule
    \end{tabular}
\end{table}
For each model we specify the number of system variables $d$, the number $d \cdot p$ of state variables in the \rodeo Kalman filter, the number of \rodeo evaluation points $N$ required to obtain a Laplace posterior indistinguishable from that of the true ODE solution. We use the interrogation of~\cite{kramer21} for the four examples. We compare our ODE solver with two deterministic solvers. The first is LSODA, which automatically switches between the Adams method for non-stiff problems and the BDF method for stiff problems~\citep{petzold83}. We use the implementation provided in the \python library \textbf{SciPy} which serves as a wrapper for the original \proglang{Fortran} solver in \textbf{ODEPACK}~\citep{hindmarsh83}. The second is RKDP discussed near the end of Section~\ref{sec:magiex} as implemented in \textbf{diffrax}.

For the univariate example, blocking is slightly slower than no blocking because there is some overhead when computing over the extra dimension. 
However, problems are typically multivariate in practice. Thus we emphasize the results of the last three examples which indicate that \rodeo is 2 to 12 times faster than RKDP and 2-12 times faster than LSODA.  We also compare timings to the $\bO(d^3)$ implementation of \rodeo which does not perform variable blocking.  Our findings indicate that blocking is 3-4 times faster. 

\newpage

\section{Discussion} \label{sec:disc}

We present \rodeo, a lightweight and extensible framework of probabilistic methods for parameter inference in ODEs, integrating methods developed by several research groups~\citep{tronarp.etal18, wu.etal24, chkrebtii.etal16, yang.etal21}. By leveraging \jax for automatic differentiation and JIT-compilation, \rodeo offers a fast and flexible toolkit for parameter inference in ODEs with partially observed components and arbitrary measurement models. The framework is compatible with several \jax-based optimization libraries, giving users the freedom to choose whichever optimizer best fits their problem. In addition to inference tools, \rodeo includes a probabilistic ODE solver that achieves performance comparable to or exceeding state-of-the-art deterministic solvers across several ODEs.

The ongoing development of the \rodeo library suggests several directions for future research. First, the \rodeo probabilistic solver currently lacks a mechanism for selecting an appropriate step size.  This could be done adaptively using techniques such as those proposed by~\cite{schober.etal19}. Second, \rodeo does not currently support parameter inference in models with time-varying parameters. Such models have been considered in earlier work~\citep[e.g.,][]{cao.etal12, meng.etal21}, and extending \rodeo to handle such models within a probabilistic framework remains an open area for development. Finally, future work could investigate how the various \rodeo inference methods perform on stiff ODE systems and boundary value problems.


\section*{Acknowledgments}

\begin{leftbar}
  This work is funded by Natural Sciences and Engineering Council of Cananda (NSERC) Discovery Grant RGPIN-2020-04364.
\end{leftbar}


\bibliography{rodeo-ref}


\appendix

\section{Kalman Recursions}\label{sec:kalmanfun}
The Kalman recursions used in Algorithms~\ref{alg:Kalman},~\ref{alg:KalmanBlock},~\ref{alg:Fenrir},~\ref{alg:dalton} and~\ref{alg:daltonng} are formulated in terms of the general Gaussian state space model
\begin{equation}
    \begin{aligned}
        \XX_n &= \QQ_n \XX_{n-1} + \cc_n + \RR_n^{1/2} \eps_n \\
        \ZZ_n &= \WW_n \XX_n + \aa_n + \VV_n^{1/2} \eet_n
    \end{aligned}, \qquad \eps_n, \eet_n \ind \N(\bz,\Id).
\end{equation}
Throughout, we use the notation $\mmu_{n|m} = \E[\XX_n \mid \ZZ_{1:m}]$ and $\SSi_{n|m} = \var(\XX_n \mid \ZZ_{1:m})$.
\begin{algorithm}[H]
    \caption{Standard Kalman filtering and smoothing functions.}\label{alg:kalman}
    \begin{algorithmic}[1]
        \Procedure{\kpredict}{$\mmu_{n-1|n-1}, \SSi_{n-1|n-1}, \cc_n, \QQ_n, \RR_n$} 
        \State $\mmu_{n|n-1} \gets \QQ_n\mmu_{n-1|n-1} + \cc_n$
        \State $\SSi_{n|n-1} \gets \QQ_n\SSi_{n-1|n-1}\QQ_n' + \RR_n$
        \State \textbf{return} $\mmu_{n|n-1}, \SSi_{n|n-1}$ 
        \EndProcedure
        \Procedure{\kupdate}{$\mmu_{n|n-1}, \SSi_{n|n-1}, \ZZ_n, \aa_n, \WW_n, \VV_n$} 
        \State $\AA_n \gets \SSi_{n|n-1} \WW_n'[\WW_n\SSi_{n|n-1} \WW_n' + \VV_n]^{-1}$
        \State $\mmu_{n|n} \gets \mmu_{n|n-1}+\AA_n(\ZZ_n - \WW_n\mmu_{n|n-1} - \aa_n)$
        \State $\SSi_{n|n} \gets \SSi_{n|n-1} - \AA_n \WW_n\SSi_{n|n-1}$
        \State \textbf{return} $\mmu_{n|n}, \SSi_{n|n}$ 
        \EndProcedure
        \Procedure{\ksmooth}{$\mmu_{n+1|N}, \SSi_{n+1|N}, \mmu_{n|n}, \SSi_{n|n}, \mmu_{n+1|n}, \SSi_{n+1|n}, \QQ_{n+1}$} 
        \State $\AA_n \gets \SSi_{n|n}\QQ_{n+1}'\SSi_{n+1|n}^{-1}$
        \State $\mmu_{n|N} \gets \mmu_{n|n} + \AA_n(\mmu_{n+1|N} - \mmu_{n+1|n})$
        \State $\SSi_{n|N} \gets \SSi_{n|n} + \AA_n(\SSi_{n+1|N} - \SSi_{n+1|n}) \AA_n'$
        \State \textbf{return} $\mmu_{n|N}, \SSi_{n|N}$ 
        \EndProcedure
        \Procedure{\ksample}{$\XX_{n+1}, \mmu_{n|n}, \SSi_{n|n}, \mmu_{n+1|n}, \SSi_{n+1|n}, \QQ_{n+1}$} 
        \State $\AA_n \gets \SSi_{n|n}\QQ_{n+1}'\SSi_{n+1|n}^{-1}$
        \State $\tilde{\mmu}_{n|N} \gets \mmu_{n|n} + \AA_n(\XX_{n+1} - \mmu_{n+1|n})$ \Comment{$\tilde{\mmu}_{n|N} = \E[\XX_n \mid \XX_{n+1:N}, \ZZ_{1:N}]$}
        \State $\tilde{\SSi}_{n|N} \gets \SSi_{n|n} - \AA_n\QQ_{n+1}\SSi_{n|n}$ \Comment{$\tilde{\SSi}_{n|N} = \var[\XX_n \mid \XX_{n+1:N}, \ZZ_{1:N}]$}
        \State $\XX_n \sim \N(\tilde{\mmu}_{n|N}, \tilde{\SSi}_{n|N})$
        \State \textbf{return} $\XX_n$
        \EndProcedure
        \Procedure{\kforecast}{$\mmu_{n|n-1}, \SSi_{n|n-1}, \aa_n, \WW_n, \VV_n$} 
        \State $\lla_{n|n-1} \gets \WW_n \mmu_{n|n-1} + \aa_n$ \Comment{$\lla_{n|n-1} = \E[\ZZ_n \mid \ZZ_{1:n-1}]$}
        \State $\OOm_{n|n-1} \gets \WW_n \mmu_{n|n-1} \WW_n' + \VV_n$ \Comment{$\OOm_{n|n-1} = \var[\ZZ_n \mid \ZZ_{1:n-1}]$}
        \State \textbf{return} $\lla_{n|n-1}, \OOm_{n|n-1}$ 
        \EndProcedure
        \Procedure{\kcond}{$\mmu_{n|n}, \SSi_{n|n}, \mmu_{n+1|n}, \SSi_{n+1|n}, \QQ_{n+1}$}
        \State $\AA_n \gets \SSi_{n|n}\QQ_{n+1}'\SSi_{n+1|n}^{-1}$
        \State $\bb_n \gets \mmu_{n|n} - \AA_n \mmu_{n+1|n}$
        \State $\CC_n \gets \SSi_{n|n} - \AA_n \QQ_n \SSi_{n|n}$
        \State \textbf{return} $\AA_n, \bb_n, \CC_n$ \Comment{$\XX_n \mid \XX_{n+1} \sim \N(\AA_n \XX_{n+1} + \bb_n, \CC_n)$}
        \EndProcedure
    \end{algorithmic}
\end{algorithm}

\section{DALTON for Non-Gaussian Measurements}\label{sec:daltonng}

In the case of the general measurement model~\eqref{eq:meas}, which we write as
\begin{equation}\label{eq:nongmeas}
    \YY_i \ind \exp\{-g_i(\YY_i \mid \xobs_{n(i)}, \pph)\},
\end{equation}
where $\xobs_{n(i)} = \DD_i \XX_{n(i)}$ and $\xun_{n(i)}$ denote observed and unobserved components of $\xx(t)$ at time $t = t_i'$, such that \mbox{$\pdv{\xun_{n(i)}} p(\YY_i \mid \xx_{n(i)}, \pph) = \bz$} and $\DD_i$ is a mask matrix of zeros and ones. In order to compute the likelihood~\eqref{eq:likepar}, omitting the dependence on $\TTh$ we consider the identity,
\begin{equation}\label{eq:nongdec}
    \begin{aligned}
        & \phantom{\;=\;} p(\YY_{0:M} \mid \ZZ_{1:N} = \bz) =  \frac{p(\YY_{0:M}, \XX_{1:N} \mid \ZZ_{1:N} = \bz)}{p(\XX_{1:N} \mid \YY_{0:M}, \ZZ_{1:N} = \bz)} \\
        & = \frac{p(\XX_{1:N} \mid \ZZ_{1:N} = \bz) \times \prod_{i=0}^M \exp\{-g_i(\YY_i \mid \xobs_{n(i)})\}}{p(\XX_{1:N} \mid \YY_{0:M}, \ZZ_{1:N} = \bz)},
    \end{aligned}
\end{equation}
which holds for any value of $\XX_{0:N}$. The numerator of~\eqref{eq:nongdec} $p(\XX_{1:N} \mid \ZZ_{1:N} = \bz)$ can be approximated using the Kalman smoothing algorithm applied to the data-free surrogate model~\eqref{eq:pmi}, whereas the product term is obtained via straightforward calculation of~\eqref{eq:nongmeas}. As for the denominator of~\eqref{eq:nongdec}, we propose to approximate it by a multivariate normal distribution which after simplifications gives the model
\begin{equation}\label{eq:pgauss}
    \begin{aligned}
        \XX_{n+1} \mid \XX_n & \sim \N(\QQ_\eet \XX_n, \RR_\eet) \\
        \ZZ_n & \ind \N(\dot \xx_n - \ff_\tth(\xx_n, t_n), \VV) \\
        \hat \YY_i & \ind \N(\xobs_{n(i)}, [\nabla^2 h_i(\hxobs_{n(i)})]^{-1}).
    \end{aligned}
\end{equation}
where $h_i(\xobs) = g_i(\YY_i \mid \xobs, \pph)$ is a short-hand notation, $\nabla h_i$ and $\nabla^2 h_i$ are the gradient and Hessian of $h_i$ respectively, and $\hat \YY_i = \hxobs_{n(i)} - \nabla^2 h_i(\hxobs_{n(i)})^{-1}\nabla h_i(\hxobs_{n(i)})$. We may now augment the surrogate model~\eqref{eq:pmi} exactly as in Gaussian setting of Section~\ref{sec:dalton} to obtain an estimate of $p_\L(\XX_{1:N} \mid \hat \YY_{0:M}, \ZZ_{1:N} = \bz)$ by the Kalman smoother. To complete the algorithm, there remains the choice $\XX_{0:N}$ to plug into~\eqref{eq:nongdec}, and the choice of $\hxobs_{n(i)}$ in~\eqref{eq:pgauss}.  For the latter, we use
$\hxobs_{n(i)} = \E_\L[\xobs_{n(i)} \mid \hat \YY_{0:i-1}, \ZZ_{1:n(i)-1} = \bz]$, 
the predicted mean of the surrogate model obtained from~\eqref{eq:pgauss}.  For the former, we use
$\XX_{0:N} = \E_\L[\XX_{0:N} \mid \hat \YY_{0:M}, \ZZ_{1:N} = \bz]$, the Kalman smoother mean of $p_\L(\XX_{1:N} \mid \hat \YY_{0:M}, \ZZ_{1:N} = \bz)$. For full details, please see~\cite{wu.etal24}.

\begin{algorithm}[!htb]
    \caption{\dalton probabilistic ODE likelihood approximation for non-Gaussian measurements.}\label{alg:daltonng}
    \begin{algorithmic}[1]
        \Procedure{\code{daltonng}}{}$\left(
        \begin{aligned}
          & \WW = \WW_\tth, \quad \ff(\XX, t) = \ff_\tth(\XX, t), \quad \vv = \vv_\tth, \\
          & \QQ = \QQ_\eet, \quad \RR = \RR_\eet, \\
          & \YY_{0:M}, \quad \DD_{0:M}, \qquad \gg_{0:M}(\YY, \XX) = \gg_{0:M}(\YY, \XX, \pph)
        \end{aligned}
      \right)$
        \State $\mmu_{0|0}, \Sigma_{0|0} \gets \vv, \bz$ \Comment{Initialization}
        \State $\ZZ_{1:N} \gets \bz$
        \State $\ell_{xz}, \ell_{xyz}, \ell_y \gets 0, 0, 0$
        \State $i \gets 0$ \Comment{Used to map $t_n$ to $t'_i$}
        \State \Comment{Lines 7-23 compute $\log p(\XX_{1:N} \mid \hat \YY_{0:M}, \ZZ_{1:N}=\bz, \TTh)$}
        \For{$n=1:N$} 
        \State $\mmu_{n|n-1}, \SSi_{n|n-1} \gets \kbpredict(\mmu_{n-1|n-1}, \SSi_{n-1|n-1}, \bz, \QQ, \RR)$
        \State $\aa_n, \BB_n, \VV_n \gets \inter(\mmu_{n|n-1}, \SSi_{n|n-1}, \WW, \ff(\XX, t_n))$
        \If{$t_n = t_{n(i)}$}
        \State $\gg_{(1)} \gets \pdv{\xobs_{n(i)}} g_i(\YY_i, \DD_i \mmu_{n|n-1})$
        \State $\gg_{(2)} \gets \pdv{}{\xobs_{n(i)}}{\xobs_{n(i)}'} g_i(\YY_i, \DD_i \mmu_{n|n-1})$
        \For{$k=1:d$}
        \State $\hat \YY_i^{(k)} \gets \DD_i^{(k)}\mmu_{n|n-1}^{(k)} - {\gg_{(2)}^{-1}}^{(k)}\gg_{(1)}^{(k)}$ \Comment{Compute pseudo-observations}
        \State $\ZZ_n ^{(k)}\gets  \begin{bmatrix}
            \ZZ_n^{(k)} \\
            \hat \YY_i^{(k)}
        \end{bmatrix}, \qquad
        \WW^{(k)} \gets \begin{bmatrix} 
            \WW \\
            \DD_i^{(k)}
        \end{bmatrix}, \qquad
        \BB_n^{(k)} \gets \begin{bmatrix}
            \BB_n^{(k)} \\
            \bz
        \end{bmatrix}$
        \State $\aa_n^{(k)} \gets \begin{bmatrix}
            \aa_n^{(k)} \\
            \bz
        \end{bmatrix}, \qquad
        \VV_n^{(k)} \gets \begin{bmatrix}
            \VV_n^{(k)} & \bz \\
            \bz & {-\gg_{(2)}^{-1}}^{(k)}
        \end{bmatrix}$
        \EndFor
        \State $i \gets i + 1$
        \EndIf
        \State $\mmu_{n|n}, \SSi_{n|n} \gets \kbupdate(\mmu_{n|n-1}, \SSi_{n|n-1}, \ZZ_n, \aa_n, \WW +\BB_n, \VV_n)$ 
        \EndFor
        \For{$n=N-1:1$} 
        
        \State $\mmu_{n|N}, \SSi_{n|N} \gets \kbsmooth(\mmu_{n+1|N}, \SSi_{n+1|N}, \mmu_{n|n}, \SSi_{n|n}, \mmu_{n+1|n}, \SSi_{n+1|n}, \QQ)$
        \State $\mmu_n, \SSi_n \gets \kbsample(\mmu_{n+1|N}, \mmu_{n|n}, \SSi_{n|n}, \mmu_{n+1|n}, \SSi_{n+1|n}, \QQ)$
        \State $\ell_{xyz} \gets \ell_{xyz} + \nlogpdf(\mmu_{n|N}; \mmu_n, \SSi_n)$
        \EndFor
        \State $\ell_{xyz} \gets \ell_{xyz} + \nlogpdf(\mmu_{N|N}; \mmu_{N|N}, \SSi_{N|N})$
        \State \Comment{Lines 25-34 compute $\log p(\XX_{1:N} \mid \ZZ_{1:N}=\bz, \TTh)$}
        \State $\ZZ_{1:N} \gets \bz$ \Comment{Reset $\ZZ_{1:N} =\bz$}
        \State $\mmu'_{0:N|N} \gets \mmu_{0:N|N}$ \Comment{Store $\mmu_{0:N|N} = \E[\XX_{0:N} \mid \ZZ_{1:N}, \hat \YY_{0:N}]$ for later}
        \For{$n=1:N$}
        \State $\mmu_{n|n-1}, \SSi_{n|n-1} \gets \kbpredict(\mmu_{n-1|n-1}, \SSi_{n-1|n-1}, \bz, \QQ, \RR)$
        \State $\aa_n, \BB_n, \VV_n \gets \inter(\mmu_{n|n-1}, \SSi_{n|n-1}, \WW, \ff(\XX, t_n))$
        \State $\mmu_{n|n}, \SSi_{n|n} \gets \kupdate(\mmu_{n|n-1}, \SSi_{n|n-1}, \ZZ_n, \aa_n, \WW + \BB_n, \VV_n$
        \EndFor
        \For{$n=N-1:1$}
        \State $\mmu_n, \SSi_n \gets \kbsample(\mmu'_{n+1|N}, \mmu_{n|n}, \SSi_{n|n}, \mmu_{n+1|n}, \SSi_{n+1|n}, \QQ)$
        \State $\ell_{xz} \gets \ell_{xz} + \nlogpdf(\mmu'_{n|N}; \mmu_n, \SSi_n)$
        \EndFor
        \State $\ell_{xz} \gets \ell_{xz} + \nlogpdf(\mmu'_{N|N}; \mmu_{N|N}, \SSi_{N|N})$
        \State \Comment{Lines 36-37 compute $\log p(\YY_{0:M} \mid \XX_{0:N}, \pph)$}
        \For{$i=0:M$} 
        \State $\ell_y \gets \ell_y + g_i(\YY_i, \DD_i\mmu'_{n(i)|N})$
        \EndFor
        \State 
        \State \textbf{return} $\ell_{xz} - \ell_y - \ell_{xyz}$ \Comment{Estimate of $\log p(\YY_{0:M} \mid \ZZ_{1:N} = \bz, \TTh)$}
        \EndProcedure
    \end{algorithmic}
\end{algorithm}

\end{document}